\documentclass[sigconf,nonacm]{acmart}
\usepackage[a-2b,mathxmp]{pdfx}[2023/10/12]
\usepackage{amsmath}
\usepackage{amsfonts}
\usepackage{amstext}
\usepackage{graphicx}
\usepackage{textcomp}
\usepackage{xcolor}
\usepackage{url}            
\usepackage{hyperref}
\usepackage{booktabs}      
\usepackage{mathrsfs}
\usepackage{graphics}
\usepackage{balance}
\usepackage{caption}
\usepackage{multirow}
\usepackage{boldline}
\usepackage{wrapfig}
\usepackage{enumitem}
\usepackage{color,colortbl}
\usepackage{soul}
\usepackage{makecell}
\usepackage{hhline}
\usepackage{subfigure}
\usepackage{wrapfig}
\usepackage{footmisc}
\usepackage{listings}
\usepackage[noend]{algpseudocode}
\usepackage[ruled, lined, linesnumbered, commentsnumbered, noend]{algorithm2e}
\usepackage{eqparbox}
\usepackage{lipsum}
\usepackage{xargs}   
\usepackage[normalem]{ulem}
\usepackage{xspace}
\usepackage{makecell}

\AtBeginDocument{%
  \providecommand\BibTeX{{%
    \normalfont B\kern-0.5em{\scshape i\kern-0.25em b}\kern-0.8em\TeX}}}

\newcommand{\ie}{\emph{i.e.}}
\newcommand{\eg}{\emph{e.g.}}
\newcommand{\etc}{\emph{etc}}

\newcommand{\cblue}[1]{{\color{blue}{#1}}}

\newcommand{\sqlq}{\mathtt{\boldsymbol{q}}\xspace}
\newcommand{\joinfeaturization}{\sqlq_{\boldsymbol{J}}\xspace}
\newcommand{\filterfeaturization}{\sqlq_{\boldsymbol{F}}\xspace}
\newcommand{\card}{{\sf c}\xspace}
\newcommand{\db}{\mathbf{D}\xspace}

\newcommand{\dbstates}{\boldsymbol{X}\xspace}
\newcommand{\adbstate}{\boldsymbol{x}\xspace}
\newcommand{\primeencout}{\boldsymbol{Z'}\xspace}
\newcommand{\encout}{\boldsymbol{Z}\xspace}
\newcommand{\anencout}{\boldsymbol{z}\xspace}
\newcommand{\answervec}{\boldsymbol{y}\xspace}
\newcommand{\nenc}{n_{\sf enc}\xspace} 
\newcommand{\nana}{n_{\sf ana}\xspace}
\newcommand{\dq}{d_q\xspace}
\newcommand{\dx}{d_x\xspace}

\newcommand\modelName{ALECE\xspace}
\newcommand\mname{ALECE\xspace}

\newcommand{\tabincell}[2]{\begin{tabular}{@{}#1@{}}#2\end{tabular}}

\newcolumntype{C}[1]{>{\centering\arraybackslash}p{#1}}
\newcolumntype{L}[1]{>{\raggedright}p{#1}}

\newlength{\oldtabcolsep}
\newlength{\oldtextfloatsep}
\newlength{\oldcolumnsep}
\newlength{\oldintextsep}
\newlength{\oldabovecaptionskip}
\newlength{\oldbelowcaptionskip}

\definecolor{red1}{RGB}{245,222,179}
\definecolor{red2}{RGB}{237,145,33}
\definecolor{red3}{RGB}{255,128,0}
\definecolor{grey}{RGB}{180,180,180}
\definecolor{green1}{HSB}{80,100,360}
\definecolor{green2}{HSB}{80,115,360}
\definecolor{green3}{HSB}{80,135,360}
\definecolor{green4}{HSB}{80,360,360}
\definecolor{aqua}{RGB}{0, 191, 255}

\definecolor{dkgreen}{rgb}{0,0.6,0}
\definecolor{gray}{rgb}{0.5,0.5,0.5}
\definecolor{mauve}{rgb}{0.58,0,0.82}

\def\releaseversion{1}
\def\includeAppendix{1}
\def\seperateAppendix{0}

\def\modifymode{0}

\if\includeAppendix1
\newcommand{\labelMultihead}{\ref{sec:multihead_attn}\xspace}
\newcommand{\labelFigJoins}{\ref{fig:joins}\xspace}
\newcommand{\labelBenchmark}{\ref{sec:benchmark_usage}\xspace}
\newcommand{\labelAdditionalExp}{\ref{sec:additional_exp}\xspace}
\else
\newcommand{\labelMultihead}{A\xspace}
\newcommand{\labelFigJoins}{5\xspace}
\newcommand{\labelBenchmark}{B\xspace}
\newcommand{\labelAdditionalExp}{C\xspace}
\fi

\if\seperateAppendix1
\newcommand{\labelQueryFeat}{3.2\xspace}
\else
\newcommand{\labelQueryFeat}{\ref{subsec:query_featurization}\xspace}
\fi

\newcommand\vldbdoi{10.14778/3626292.3626302}
\newcommand\vldbpages{197 - 210}
\newcommand\vldbvolume{17}
\newcommand\vldbissue{2}
\newcommand\vldbyear{2023}
\newcommand\vldbauthors{Pengfei Li, Wenqing Wei, Rong Zhu, Bolin Ding, Jingren Zhou, and Hua Lu}
\newcommand\vldbtitle{\modelName: An Attention-based Learned Cardinality Estimator for SPJ Queries on Dynamic Workloads} 
\newcommand\vldbavailabilityurl{https://github.com/pfl-cs/ALECE}
\newcommand\vldbpagestyle{empty} 

\begin{document}
\if\includeAppendix1
\title{\modelName: An Attention-based Learned Cardinality Estimator for SPJ Queries on Dynamic Workloads (Extended)}
\else
\title{\modelName: An Attention-based Learned Cardinality Estimator for SPJ Queries on Dynamic Workloads}
\fi

%
%
%
%
%
%

	\author{Pengfei Li}
	\affiliation{
		\institution{Alibaba Group, China}
		\country{}
	}
	\email{lpf367135@alibaba-inc.com}
	
	\author{Wenqing Wei}
	\affiliation{
		\institution{Alibaba Group, China}
		\country{}
	}
	\email{weiwenqing.wwq@alibaba-inc.com}
	
	\author{Rong Zhu $\dagger$} 
	\affiliation{
		\institution{Alibaba Group, China}
		\country{}
	}
	\email{red.zr@alibaba-inc.com}
	
	\author{Bolin Ding $\dagger$} 
	\affiliation{
		\institution{Alibaba Group, China}
		\country{}
	}
	\email{bolin.ding@alibaba-inc.com}
	
	\author{Jingren Zhou $\dagger$} 
	\affiliation{
		\institution{Alibaba Group, China}
		\country{}
	}
	\email{jingren.zhou@alibaba-inc.com}
	
	\author{Hua Lu $\dagger$} 
	\affiliation{
		\institution{Roskilde University, Denmark}
		\country{}
	}
	\email{luhua@ruc.dk}



\begin{abstract}
For efficient query processing, DBMS query optimizers have for decades relied on delicate cardinality estimation methods.
%
%
In this work, we propose an Attention-based LEarned Cardinality Estimator (\textbf{\modelName}~for short) for SPJ queries.
The core idea is to discover the implicit relationships between queries and underlying dynamic data using attention mechanisms in~\modelName's two modules that are built on top of carefully designed featurizations for data and queries.
In particular, from all attributes in the database, the data-encoder module obtains organic and learnable aggregations which implicitly represent correlations among the attributes, whereas the query-analyzer module builds a bridge between the query featurizations and the data aggregations to predict the query's cardinality.
We experimentally evaluate~\modelName~on multiple dynamic workloads. The results show that~\modelName~enables PostgreSQL's optimizer to achieve nearly optimal performance, clearly outperforming its built-in cardinality estimator and other alternatives.

%
%
\end{abstract}


\maketitle
\vspace{-1pt}
\pagestyle{\vldbpagestyle}
\begingroup\small\noindent\raggedright\textbf{PVLDB Reference Format:}\\
\vldbauthors. \vldbtitle. PVLDB, \vldbvolume(\vldbissue): \vldbpages, \vldbyear.\\
\href{https://doi.org/\vldbdoi}{doi:\vldbdoi}
\endgroup
\vspace{-1pt}
\begingroup
\renewcommand\thefootnote{}\footnote{
	\noindent \hspace{-16pt}$\dagger$ Corresponding authors. 
	
	\noindent\rule{.475\textwidth}{0.4pt}
	
	\noindent
	This work is licensed under the Creative Commons BY-NC-ND 4.0 International License. Visit \url{https://creativecommons.org/licenses/by-nc-nd/4.0/} to view a copy of this license. For any use beyond those covered by this license, obtain permission by emailing \href{mailto:info@vldb.org}{info@vldb.org}. Copyright is held by the owner/author(s). Publication rights licensed to the VLDB Endowment. \\
	\raggedright Proceedings of the VLDB Endowment, Vol. \vldbvolume, No. \vldbissue\ %
	ISSN 2150-8097. \\
	\href{https://doi.org/\vldbdoi}{doi:\vldbdoi} \\
}\addtocounter{footnote}{-1}\endgroup

\ifdefempty{\vldbavailabilityurl}{}{
	\vspace{.3cm}
	\begingroup\small\noindent\raggedright\textbf{PVLDB Artifact Availability:}\\
	The source code, data, and/or other artifacts have been made available at \url{\vldbavailabilityurl}.
	\endgroup
}

\section{Introduction}
\label{sec:intro}

A cardinality estimator in a DBMS~\cite{DBLP:conf/sigmod/SelingerACLP79,DBLP:conf/icde/GraefeM93} estimates the number of result elements of a SQL query before query execution, and thus helps the query optimizer to generate good query plans.
In the past, the mainstream of cardinality estimation has always been statistical data-driven methods. Such methods condense information about data into lightweight summaries, \emph{e.g.,} histograms, sketches and data distribution approximation, and adopt analytic functions with the summaries as the input to estimate cardinatilies of SQL queries~\cite{DBLP:conf/sigmod/SelingerACLP79,DBLP:journals/tods/KaushikNRC05,DBLP:journals/pvldb/ZhuWHZPQZC21}. However, real-world datasets are often complex and the analytic functions are usually not powerful enough to build correct mappings between coarse data summaries and SQL query cardinalities.
Also, SQL queries often contain join predicates but it is difficult and time-consuming to build particular summaries for each join. Computing joint data distributions is also usually intractable due to high computation and storage overhead.

Recently, traditional cardinality estimators have been disrupted by estimators based on learned models. 
Data-driven models~\cite{DBLP:journals/pvldb/YangLKWDCAHKS19,DBLP:journals/pvldb/HilprechtSKMKB20,DBLP:journals/pvldb/YangKLLDCS20,DBLP:journals/pvldb/WangCLL21} learn tighter data distributions from the underlying database and use analytic expressions to estimate the cardinalities. 
In contrast, query-driven models~\cite{DBLP:conf/cidr/KipfKRLBK19,DBLP:conf/sigmod/ZhaoYHLZ22} utilize the feedback of executed queries in a supervised fashion. The latter learn the relation between cardinalities and query distributions, without paying particular attention to the underlying database.
However, neither kind of models can fully make use of both data and queries. 
It is difficult for them to extract individualized useful information for different queries.
A few models~\cite{DBLP:journals/pvldb/DuttWNKNC19,DBLP:conf/cidr/KipfKRLBK19,DBLP:journals/pvldb/NegiWKTMMKA23} consider both data and queries. However, they either only use simple and trivial data information and requires sampling operations over relations~\cite{DBLP:conf/cidr/KipfKRLBK19, DBLP:conf/sigmod/WuC21}, or do not support processing queries with joins~\cite{DBLP:journals/pvldb/DuttWNKNC19} or complex joins~\cite{DBLP:journals/pvldb/NegiWKTMMKA23}.

In addition, existing models have a more critical problem: They do not perform well on \textbf{dynamic workloads that mix queries and data manipulation statements including inserts, deletes and updates.}
Such statements tend to make estimations difficult as they influence the data distribution and shift the mapping between true cardinalities and query distributions.
When the underlying data changes, the joint data distributions among relations and attributes as well as the mapping between queries and true cardinalities also become different. Thus, pure data- or query-driven methods can hardly work on dynamic workloads.
Some methods~\cite{DBLP:conf/cidr/KipfKRLBK19, DBLP:conf/sigmod/WuC21}, although they consider both data and queries, will also have degraded performance on dynamic workloads as their required featurization or sampling approaches does not support model training and inference with data updates.
More importantly, existing methods do not answer \textbf{how to reasonably link SQL queries and the underlying data} and build an appropriate mapping among the true cardinalities, queries and data---especially when data is dynamic.

\begin{figure}[!htb]
	\centering
	\vspace{-14pt}
	\if\includeAppendix1
	\includegraphics[width=0.49\textwidth, trim={5mm 3mm 3mm 5mm},clip]{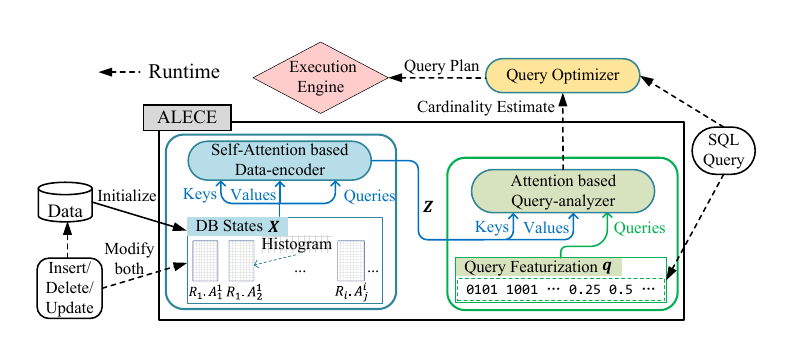}
	\else
	\includegraphics[width=0.49\textwidth, trim={5mm 3mm 3mm 5mm},clip]{overview.eps}
	\fi
	\vspace{-25pt}
	\caption{\modelName-based Query Execution.}
	\label{fig:overview}
	\vspace{-18pt}
\end{figure}

To address these drawbacks, we design an Attention-based LEarned Cardinality Estimator (\modelName) for select-project-join (SPJ) queries.
Fig.~\ref{fig:overview} depicts~\modelName~in the context of DBMS's query execution.
%
%
\modelName~is both data- and query-driven.
When estimating an SPJ query's cardinality, it losslessly featurizes the query into a vector. 
Meanwhile, it efficiently featurizes the current underlying data in the database into a set of vectors, called \textbf{\textit{DB states}}, which `compress' the whole database. Both query and data featurizations are of low space overhead and can be efficiently computed.
On top of the DB states and query featurizations,~\modelName~builds a neural network based model to create reasonable connections between them. 
The model integrates the information of the DB states and query featurization, and feeds them into a feed-forward regression neural network to make the estimates.
%
%
Roughly speaking,~\modelName~first learns to assign different weights to the raw DB states $\dbstates$, with each weight showing the correlation between two elements in $\dbstates$. This correlation is an useful distribution information to build suitable mapping between the caridinality of a SQL query with the underlying data.
Then, $\dbstates$ is mapped into another set of vectors $\encout$ which are the weighted combinations of $\dbstates$ and better represent the underlying data.
~\modelName~also learns an another weight for each mapped vector $\anencout_i$ in $\encout$ and the query featurization $\sqlq$ to measure the influence of $\anencout_i$ on $\sqlq$.
The weighted combination of $\encout$ is a convolution of the DB states and the query featurization. 
The combination vector is finally used to generate the cardinality estimate.

Our~\modelName's design encounters two challenges. First, we need to build the DB states suitable for dealing with data changes in dynamic workloads, and meanwhile make them efficient to access. 
To this end, we propose a simple yet effective data featurization approach that is a good approximation of data distributions, sensitive to data changes, and computation-efficient. 
This approach constructs succinct summaries of the underlying data, \emph{i.e.}, DB states, based on the histogram of each attribute of the database relations.
Each time a record is inserted, deleted or updated, we only need to modify the DB states' vectors relevant to the changed relation. Also, the basic single attribute distributions and even joint distributions are covered by the DB states. 
These factors together enable us to process dynamic workloads with the DB states.
Besides, depending on the requirement of the distribution approximation precision, the number of bins in a histogram can be flexibly adapted. Moreover, other useful information relevant to underlying data can be seamlessly integrated if needed.

Second, we need to extract the implicit relevance between SQL queries and corresponding DB states, and make the information helpful for cardinality estimation.
To this end, we adopt the attention mechanism~\cite{DBLP:journals/corr/BahdanauCB14,DBLP:conf/nips/VaswaniSPUJGKP17,DBLP:conf/iclr/KimDHR17} in our model to draw global dependencies between SQL queries and underlying data. The attention mechanism is widely used in a variety of tasks including question answering~\cite{DBLP:conf/nips/HermannKGEKSB15,DBLP:conf/nips/SukhbaatarSWF15}. 
Generally, it simulates the process of selection from a set using an attention function that takes as input two main components: a set of queries and a set of key-value pairs. It figures out in an individualized manner which parts of the data play more important roles for different queries, assigns higher weights to the more important and relevant keys for each query, and outputs the combination of the weighted values. 
Unlike those concepts in a database, a query in attentions is a specific element for which we need to learn a representation, the role of keys is to respond more or less to the query, and the values are used to compose an answer.
Nevertheless, the selection process exactly matches our settings where the SQL queries and underlying data are analogies of the queries and key-value pairs in attentions, respectively.

There are two modules in our~\modelName~where the attentions are used in different ways. On the one hand, the `data-encoder' module uses a self-attention whose inputs of queries, keys and values all come from the DB states. The self-attention allows the DB states, which correspond to different attributes, to interact with each other. By using the self-attention, the data-encoder module learns the implicit joint distribution information among attributes and computes a smarter representation of the underlying data.
On the other hand, in the `query-analyzer' module, the queries set of the attention is exactly a set covering only one featurization vector of a SQL query, while the keys and values come from the output of the data-encoder module. The query-analyzer module outputs a fixed-dimensional `answering' vector integrating the information from the query and data representations. We then use a simple linear regression model to map the answering vector to a cardinality estimate.

Compared to the state-of-the-art cardinality estimation methods,~\modelName~is able to make more reasonable use of both queries and underlying data. 
With the help of the two attentions, it answers the questions that `which parts of data should a SQL query pays more attention to?' and `how to find the more important data?' 
A SQL query usually focuses on some local parts of selected attributes. Also, the join conditions make particular tuples contribute more to the cardinality. 
Moreover,~\modelName~is able to adapt to dynamic workloads. In practice, learned models need to be trained with past queries and corresponding DB states, and estimate cardinalities for future queries. The performance of existing query-driven models often dramatically degrades when making predictions on a dynamic database. 
In contrast,~\modelName~can make immediate and suitable reactions to data changes by modifying the DB states, and learn an appropriate but implicit mapping between the true cardinality and the query featurization accompanied with the corresponding DB states. Our experimental results show that~\modelName~is able to make accurate estimates even when the distribution of the underlying data changes. Thus,~\modelName~is less sensitive to data changes.

In our evaluation,~\modelName~achieves the best cardinality estimation performance on multiple dynamic workloads. 
Experimental results show~\modelName~improves the average end-to-end query time by up to $2.7\times$ faster on the benchmark workload, very close to the optimal results acquired by using true cardinalities. This demonstrates that our~\modelName~makes more accurate cardinality estimates and helps the query optimizer find better query plans.

We make the following major contributions in this paper:
\vspace{-2pt}
\begin{itemize}[leftmargin=*]
	\item We propose necessary principles for a method to featurize the underlying database data and SPJ queries. Accordingly, we design a featurization schema to losslessly featurize an SPJ query and make a reasonable compression of the data. The featurizations can be efficiently updated to support dynamic workloads.
	
	\item Based on the featurizations of queries and data, we propose an attention based learned cardinality estimator~\modelName, together with detailed analyses.
	
	\item \modelName~is designed to be a `whitebox' which not only gives estimates but also clear rationale to integrate the SPJ queries and underlying data together in processing dynamic workloads.
	
	\item We experimentally validate~\modelName’s advantages over more than half dozen representative alternatives on real datasets.
\end{itemize}
\vspace{-2pt}

The rest of the paper is organized as follows. Section~\ref{sec:preliminaries} gives the preliminaries. 
Section~\ref{sec:featurizations} presents the featurizations of data and queries. 
Section~\ref{sec:approach} elaborates on~\modelName, followed by an analysis of it in Section~\ref{sec:analysis}.
Section~\ref{sec:experiments} reports on the experimental studies.
Section~\ref{sec:related_work} reviews the related work. 
Section~\ref{sec:conclusion} concludes the paper.
\if\includeAppendix1
In addition, due to space limit, we introduce our developed benchmark, which integrates \modelName into PostgreSQL's query optimizer, and more experimental analyses in~\hyperref[Appendix]{Appendix}.
\else
In addition, due to space limit, we introduce our developed benchmark, which integrates \modelName into PostgreSQL's query optimizer, and more experimental analyses in an extended version~\cite{extended_url}.
\fi

\vspace{-3pt}
\section{Preliminaries and Problem}
\label{sec:preliminaries}
\vspace{-2pt}
%
%

Table~\ref{tab:notations} lists important notations used in the paper.

\begin{scriptsize}
\begin{table}[htb]
    \centering
    \vspace{-10pt}
    \caption{Notations}\label{tab:notations}
    \vspace{-10pt}
    \setlength\tabcolsep{1.8pt}
    \setlength{\textfloatsep}{2pt}
    \begin{tabular}{l|l} \hline 
	$R_i$ & A relation in the database \\ \hline
	$A^{i}_j$ & The $j$th attribute of the relation $R_i$ \\ \hline
     $N$, $T$ & The number of relations/attributes in the database \\ \hline 
	$\dbstates = \{\adbstate_i\}_{i=1}^T$ & The set of data featurizations (a.k.a. DB states) \\ \hline
	$\dx$ & The number of histogram bins (dimensionality) for a DB state \\ \hline
	$\sqlq = \langle \joinfeaturization, \filterfeaturization \rangle$ & A SQL query and its vectorized featurization \\ \hline
	$\dq$ & The dimension of a query featurization vector \\ \hline
   	$\nenc$, $\nana$ & The number of attention layers in the data-encoder/query-analyzer module \\ \hline
        $\boldsymbol{K}, \boldsymbol{V}, \boldsymbol{Q}$ & The input keys/values/queries of an attention function \\ \hline
    \end{tabular}
 \vspace{-16pt}
\end{table}
\end{scriptsize}

\subsection{Cardinality Estimation Problem}
\label{subsec:research_problem}
\vspace{-3pt}

\if\modifymode1
\pf{I have reorganized the structure of this subsection. The problem formulation is given in advance, before introducing how~\modelName~supports the optimization of SPJ queries and even more complicated queries.}
\fi

Suppose a database $\db$ has a set of relations $\{R_1, \cdots, R_N\}$. A relation $R_i$ has $n_{i}$ attributes, \ie, $R_i = (A_{1}^{i},\cdots,A_{n_i}^{i})$. Each attribute $A_{j}^{i}$ can be either {\em categorical} or {\em numerical}: the domain of a categorical attribute is a finite set and can be 1-to-1 mapped to an integer set $\{1, \cdots, {\sf max}^i_{j}\}$; the domain of a numerical one is $[{\sf min}^i_{j} ,{\sf max}^i_{j}]$. 
%

\noindent\textbf{\underline{Problem Formulation.}} Given a SQL query $\sqlq$ and a dynamic database $\db$, we want to estimate the {\em cardinality of $\sqlq$}, denoted as $\card(\sqlq, \db)$, \ie, the number of resulting tuples when $\sqlq$ is executed on $\db$. 
 
In this paper, we focus on select-project-join SQL queries with conjunctive filter predicates; the cardinality $\card(\sqlq, \db)$ is the number of tuples after joins and filters, as the following counting query:
\vspace{-2pt}
\begin{flalign} \label{Eq:sub_query_format}
	\card(\sqlq, \db):~&\mathtt{SELECT~}\mathtt{COUNT}(*)\mathtt{~FROM~}R_{i_1},\cdots,R_{i_n} \\ \nonumber
	&\mathtt{WHERE~}\mathtt{join~predicates~} \boldsymbol{J}~\mathtt{AND~filter~predicates~} \boldsymbol{F}
	\vspace{-10pt}
\end{flalign}
\vspace{-12pt}

\noindent where $\sqlq$ involves $n$ relations $R_{i_1},\cdots,R_{i_n}$, with a set of join predicates $\boldsymbol{J}$ which is a conjunction of join conditions each in the form of ``$R_i.A_{x}^{i} = R_j.A_{y}^{j}$'', and a conjunction of filter predicates $\boldsymbol{F}$. This formulation allows us to support not only PK-FK joins but also more general joins by specifying join predicates on pairs of joinable attributes (which may or may not be primary/foreign keys) in $\boldsymbol{J}$.
%
%
%
A filter predicate is an relational expression in the form of ``$R_i.A_{j}^{i}~\mathtt{op}~{\sf const}$'' where $\mathtt{op} \in \{<, \leq, >, \geq, =\}$ and ${\sf const}$ is a fixed value. 
In $\sqlq$, an attribute can appear in a join or a filter predicate, or both. 
The support for $\mathtt{LIKE}$ predicates is left for future work.
%

\noindent\textbf{\underline{\modelName~in the Optimization of SPJ Queries.}}
The estimation results for counting queries in the format of \eqref{Eq:sub_query_format} can be used to support the optimization of more complex queries, \eg, widely-used \textbf{SPJ} (select-project-join) queries 
in the following format:
	\vspace{-2pt}
	\begin{flalign} \label{Eq:query_format}
		\sqlq_{\sf SPJ}:~&\mathtt{SELECT~}\mathtt{AGG}_1, \cdots \mathtt{AGG}_m \mathtt{~FROM~} {R_{i_1},\cdots,R_{i_n}} \\ \nonumber
		&\mathtt{WHERE~}\mathtt{join~predicates~} \boldsymbol{J}~\mathtt{AND~filter~predicates~} \boldsymbol{F} \\ \nonumber
		&\mathtt{ORDER~BY~} \mathtt{attribute\_set\_1} \mathtt{~~~GROUP~BY~} \mathtt{attribute\_set\_2}
		\vspace{-8pt}
	\end{flalign}
\vspace{-12pt}

\noindent where each $\mathtt{AGG}_k$ ($k = 1, \cdots, m$) is an aggregate function over one or multiple attributes which can be $\mathtt{COUNT}$, $\mathtt{AVG}$, $\mathtt{MIN}$ and $\mathtt{MAX}$, \emph{etc}, or can be simply omitted. The join predicate set $\boldsymbol{J}$ and filter predicate set $\boldsymbol{F}$ carry the same meanings with that in \eqref{Eq:sub_query_format}.

To search for the best execution plan, the query optimizer of a modern DBMS like PostgreSQL first decomposes $\sqlq_{\sf SPJ}$ into a series of sub-queries (implicitly) in some fixed order~\cite{DBLP:journals/pvldb/HanWWZYTZCQPQZL21}.
The cardinalities of these sub-queries are then estimated with the built-in estimator. Accordingly, candidate query execution plans are enumerated and their estimated execution costs given the cardinality estimates are calculated using also a fixed cost model. The plan with the smallest estimated cost is chosen to execute the query physically. Apparently, the execution performance of a query $\sqlq_{\sf SPJ}$ is basically determined by the cardinality estimates of its sub-queries.
The design of our \modelName enables it to provide more accurate estimates. Our developed benchmark can plug external cardinality estimator into the optimizer to replace the built-in one. Thus, \modelName is able to improve the cardinality estimation for $\sqlq_{\sf SPJ}$'s sub-queries, and further enable the query optimizer to select a good execution plan. 

\modelName is applicable for optimizing even more complex queries. For the sub-queries that it supports, it gives better cardinality estimates; for the sub-queries that \modelName does not support, the optimizer can still use its default cardinality estimator.

%



\noindent\textbf{\underline{Estimation Model on Dynamic Workloads.}}
In reality, the data in a DBMS is seldom static but often keeps being updated. Thus, it is beneficial to design cardinality estimators able to make accurate estimates for a {\em dynamic workload}, \ie, a sequence of SQL statements including queries, inserts, deletes and updates. 
Our \modelName~ {\em aims to support dynamic workloads and provide up-to-date cardinality estimates for queries at any time during the workload}. 

For a learned cardinality estimator to work on frequent changes of the underlying data distribution without retraining, a straightforward idea is to use the database $\db$ itself as part of the input features to train an estimation model. However, this is infeasible as the size of $\db$ can be huge and varies continuously. Instead, we use succinct summaries of the database (\eg, fixed-size histograms), called {\em DB states}, as part of input features to train \modelName.
As the database $\db$ is updated, the DB states should be updated accordingly (and efficiently) such that they can be fed into the trained model to produce cardinality estimates.
%
%
%
Details about featurizing $\db$ as DB states are in Section~\ref{subsec:data_featurization}.
We assume that $\db$'s schema is static. The support for dynamic schema is left for future work.

\vspace{-2pt}
\subsection{Overview of~\modelName}
\label{subsec:overview}
\vspace{-1pt}

An overview of \modelName's model structure and its role in the query engine is shown in Fig.~\ref{fig:overview}.
Features from dynamic data and queries are decoupled and handled by two modules in \modelName. The {\em data-encoder} module adopts a self-attention structure on DB states to figure out the correlations among all attributes and to learn their joint distribution, whereas the {\em query-analyzer} module employs a data-query cross attention to discover correlation between the data-encoder's outputs and the cardinalities of (sub-)queries.
%
%
The cardinality estimates eventually produced by the query-analyzer in \modelName depend on both data (DB states) and queries. 
%

\noindent\underline{\textbf{Offline Training.}} Training~\modelName~needs a dataset of queries, their true cardinalities, and the corresponding database information when these queries are executed. 
The training dataset is obtained by collecting the true cardinalities of the historical queries executed on a dynamic database for a period of time.
We start from featurizing the initial database by generating a set of vectors with fixed dimensionality, \emph{i.e.}, DB states (Section~\ref{subsec:data_featurization}). Statements in the SQL workload are sequentially processed. For the insert, delete and update statements, we modify the DB states accordingly. 
When a query comes, it will also be featurized into a vector with fixed dimensionality (Section~\ref{subsec:query_featurization}). The query features and the current DB states will be packed together as a training sample with the true cardinality as its label. With sufficient training samples collected, \modelName is trained with gradient descent methods.

\noindent\underline{\textbf{Online Estimation.}} A well-trained \modelName can make online estimates on both static and dynamic workloads. 
%
%
Given a new query, we feed its featurization and the up-to-date DB states into \modelName to get the estimates. If the workload is static, \ie, it contains no data update statements, the DB states are constant. Otherwise, the DB states keep changing and the latest ones will always be used.
%
\section{Featurizations of Data and Queries}
\label{sec:featurizations}

The underlying database data and SQL queries are required to be featurized numerically such that our~\modelName~can deal with.
Any featurization method is able to be flexibly adopted by~\modelName~as long as it satisfies some principles.
First, the underlying data needs to be featurized into a set of fixed-dimensional vectors covering enough distribution information.
Second, any SPJ query should be losslessly mapped to a fixed-dimensional vector such that~\modelName~could better understand it.
Also, to effectively support the dynamic workloads, the featurization method is supposed to be efficient and of low storage overhead.
Following these principles, we propose an method of featurizing the database data and SQL queries numerically.
The details are given in Section~\ref{subsec:data_featurization} and~\ref{subsec:query_featurization}, respectively. Moreover, 
Section~\ref{subsec:featurization_discussion} discusses the properties of our featurization method and how it helps process dynamic workloads.
It is noteworthy that our featurization method is specifically designed for our \modelName and it perfectly aligns with the requirements on inputs to~\modelName.

\vspace{-5pt}
\subsection{Data Featurization}
\label{subsec:data_featurization}
\vspace{-2pt}
In our settings, the data featurization $\dbstates$, also known as the `DB states', is a compression of the whole database, which can roughly describe the data of each attribute and the relationships among them. 
Our~\modelName~requires the data featurization to be a set of vectors of the same dimension.
Here we use the set of histograms for each attribute as the DB states, \emph{i.e.,} $\dbstates = \{\adbstate_1, \cdots, \adbstate_T\}$ where $\adbstate_i$ is the histogram of the $i$th attribute and $T = \sum_{i = 1}^{N} n_i$ is the number of all attributes in the database.
How to order the $T$ attributes will be introduced in Section~\ref{subsec:query_featurization}.
This featurization method is simple but powerful and we can efficiently access and update the histograms.

In particular, the values of categorical attributes are first converted to consecutive integers numbered from 1. Given an attribute $A_i$, we use $\mathsf{dom}(A_i)$ to denote its domain or the converted integer set if $A_i$ is categorical. 
It is easy to show that $\mathsf{dom}(A_i) \subseteq \mathsf{D}(A_i) = [l, u)$ where $l = \inf \mathsf{Dom}(A_i)$ and $u = \sup \mathsf{Dom}(A_i) + \epsilon$ with $\epsilon$$\rightarrow$$0^{+}$. 
Then, given a time stamp $t$ and the database data at $t$, we create a $\boldsymbol{\dx}$-bin-histogram for each $A_i$. In particular, let $a = \frac{u - l}{\dx}$ and $\beta_j$ be the number of $A_i$'s values in $[l + (j - 1) \cdot a, l + j \cdot a)$ for $1 \leq j \leq \dx$, the histogram $\adbstate_i$ for attribute $A_i$ could be easily accessed with $\adbstate_i = [\beta_1, \cdots, \beta_{\dx}]$.
Our DB states at any time is simply a set of $T$ elements each being a $\dx$-dim histogram vector.
In practice, each $\beta_i$ in all histogram vectors will be scaled to the range [0, 1] through a suitable affine transformation. The value of $\dx$ can be flexibly modified according to the complexity of the data distribution. 
Apparently, a larger $\dx$ will make the data featurization capture more distribution information among the attributes but result in extra time and storage overhead.
Usually, when strong correlations exist among attributes, a larger value tends to be used for $\dx$.

\vspace{-4pt}
\subsection{Query Featurization}
\label{subsec:query_featurization}
\vspace{-1pt}

Following and extending the existing work~\cite{DBLP:conf/sigmod/ZhaoYHLZ22,DBLP:journals/pvldb/WangQWWZ21}, we featurize a SQL query $\sqlq$ into a fixed length vector $\sqlq$\footnote{Without ambiguity, $\sqlq$ denotes both a SQL query and its vectorized featurization.}.
It is a simple concatenation of two separately generated parts $\joinfeaturization$ and $\filterfeaturization$, which featurize the join predicates $\boldsymbol{J}$ and filter predicates $\boldsymbol{F}$, respectively.

\noindent\textbf{\underline{Join featurization.}} 
For the $N$ relations numbered from 1 to $N$, 
we use $m_1 = \lceil \log_{2} (N+1) \rceil$ bits to featurize the id of each relation. 
Similarly, $m_2 = \lceil \log_{2} (n_{max}+1) \rceil$ bits can featurize the ids of all attributes in any relation, where $n_{max} = \max(\{n_1, \cdots, n_N\})$ is the maximum number of attributes in a relation. 
Thus, any attribute $R_i.A^{i}_j$ can be uniquely identified with a binary vector of dimension $m = m_1 + m_2$. The first $m_1$-dimensional and the last $m_2$-dimensional sub-vectors identify the relation and the attribute, respectively. Then a join predicate $P$ is featurized by a $2m$-dim binary vector $E_J(P)$ with the first and second half sub-vectors refer to the left and right hand side of $P$, respectively.

Suppose there are $\Delta$ possible join patterns, the join featurization $\joinfeaturization$ of a SQL query $\sqlq$ is a $(2m \cdot \Delta)$-dim binary vector, containing $\Delta$ $2m$-dim sub-vectors, indicating which join patterns $\sqlq$ covers and featurizing their referred attributes. If the join predicates of $\sqlq$ contain the $i$th join pattern $P_i$, the $i$th sub-vector of $\joinfeaturization$ equals $E_J(P_i)$. Otherwise, this sub-vector is set to zero. 
Usually, the value of $\Delta$ is not large, nearly linearly related to the number of attributes~\cite{DBLP:journals/pacmmod/WuNAKM23,DBLP:journals/pvldb/ZhuWHZPQZC21}.
Unlike existing work~\cite{DBLP:conf/sigmod/ZhaoYHLZ22} that simply featurizes whether each join condition appears in the query, our featurization way is more compact and incorporates more helpful information about the joins.

In practice, we define the attributes in the left and right hand side of a join predicate are `equivalent' and find all equivalence classes in $\boldsymbol{J}$. Afterwords, we re-organize the join predicates based on the equivalence classes. Given an equivalence class $C = [A_1, \cdots, A_t]$ containing $t$ attributes sorted by their $m$-dim featurizations, we re-create $(t-1)$ join predicates with the $i$th one to be $A_i = A_{i+1}$. By doing this to each equivalence class and packing the corresponding join predicates together, we generate a new join predicate set $\boldsymbol{J'}$. The join featurization is actually performed with $\boldsymbol{J'}$ instead of $\boldsymbol{J}$.
In this way, two equivalent join predicate sets in explicitly different forms will be featurized to be the same. For example, in the following formula, join predicate set $\boldsymbol{J_1}$ and $\boldsymbol{J_2}$ both have two same equivalence classes. They will be converted to another set $\boldsymbol{J'}$. 

	\vspace{-10pt}
\begin{small}
\begin{flalign} \nonumber
\centering
\begin{alignedat}{4}
\boldsymbol{J_1}:&~\big(A^{1}_1 = A^{2}_1 \text{~and~} A^{2}_1 = A^{3}_1 \big) &&\text{~and~} &&\big(A^{3}_3 = A^{2}_2\big)&& \\ \nonumber
\boldsymbol{J_2}:&~\big(A^{1}_1 = A^{2}_1 \text{~and~} A^{3}_1 = A^{2}_1 \big) &&\text{~and~} &&\big(A^{2}_2 = A^{3}_3\big)&& \\ \nonumber
\boldsymbol{J'}:&~\underbrace{\big(A^{1}_1 = A^{2}_1 \text{~and~} A^{2}_1 = A^{3}_1 \big)}_{\text{Equi-class-1}} &&\text{~and~} &&\underbrace{\big(A^{2}_2 = A^{3}_3\big)}_{\text{Equi-class-2}}&&
\end{alignedat}
\end{flalign}
\end{small}
	\vspace{-8pt}

\noindent\textbf{\underline{Filter featurization.}} We sort all attributes according to their $m$-dim featurizations and use $A_k$ to denote the $k$th one among $T$ attributes.
Without loss of generality, we assume $\mathsf{D}(A_{k})$ is $[0,1)$ for each attribute $A_k$. Thus, {the product space $\mathsf{D}(A_{1}) \times \cdots \mathsf{D}(A_{k}) \times \cdots \times \mathsf{D}(A_T) = [0,1)^{T}$}. Apparently, the filter predicates $\boldsymbol{F}$ are equivalent to a hyper-rectangle $[l_1, u_1) \times \cdots [l_T, u_T)$ which is a subset of $[0,1)^{T}$.
In particular, for any filter condition on the attribute $A_k$, we convert it into an equivalent one in the form like $\sigma_{lb \leq A_k < ub}$. Then,  the values of $l_k$ and $u_k$ are set to $lb$ and $ub$, respectively.
Specifically, 

	\vspace{-11pt}
\begin{small}
\begin{flalign} \nonumber
	&(lb \leq A_k) \sim (lb \leq A_k < 1), ~~~(lb < A_k) \sim (lb - \epsilon \leq A_k < 1), \\ \nonumber 
	&(A_k < ub) \sim (0 \leq A_k < ub), ~~~(A_k \leq ub) \sim (0 \leq A_k < ub + \epsilon)\\ \nonumber
	& (A_k = x) \sim (x \leq A_k < x + \epsilon), \text{~where~} \epsilon \rightarrow 0^{+}.
\end{flalign}
\end{small}
	\vspace{-13pt}

\noindent Above, $\sim$ denotes the equivalence operator.

Accordingly, the featurization of filter predicates $\filterfeaturization$ is a $2T$-dim vector composed of the boundary points of the search hyper-rectangle, \emph{i.e.,} $E_f = [l_1, u_1, l_2, u_2, \cdots, l_T, u_T]$. In practice, each $l_i$ and $u_i$ will be normalized to [0, 1].

\noindent\textbf{\underline{Concatenation.}} By concatenating $\joinfeaturization$ and $\filterfeaturization$, we get the $\dq$-dim featurization vector of SQL query $\sqlq$.
Fig.~\ref{fig:query_featurization_example} shows an example.

\setlength{\textfloatsep}{1pt}
\setlength{\columnsep}{1pt}
\setlength{\intextsep}{1pt}
\begin{small}
\begin{figure}[htb]
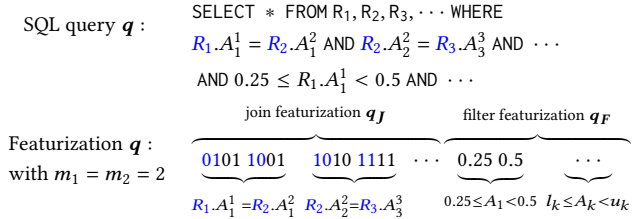

	\vspace{2pt}
	\centering
	\setcellgapes{2pt}
	\makegapedcells
	\begin{tabular}{r l}
		\multirow{2}{*}{$\text{SQL query}~\sqlq:~$} & 
		$\mathtt{SELECT~*~FROM~R_1, R_2, R_3, \cdots} \mathtt{~WHERE~}$ \\ 
		& $\cblue{R_1}.A^{1}_1 = \cblue{R_2}.A^{2}_1 \mathtt{~AND~} \cblue{R_2}.A^{2}_2 = \cblue{R_3}.A^{3}_3 \mathtt{~AND~} \cdots$ \\
		& $\mathtt{~AND~} 0.25 \leq R_1.A^{1}_1 < 0.5 \mathtt{~AND~} \cdots$ \\ 
		\tabincell{l}{$\text{Featurization}~\sqlq:$\\$\text{with~} m_1 = m_2 = 2$} & $\overbrace{\underbrace{\cblue{01} 01~\cblue{10}01}_{\cblue{R_1}.A^{1}_1\ = \cblue{R_2}.A^{2}_1}~~\underbrace{\cblue{10}10~\cblue{11}11}_{\cblue{R_2}.A^{2}_2 = \cblue{R_3}.A^{3}_3}~~\cdots}^{\text{join featurization~} \joinfeaturization} \overbrace{\underbrace{0.25~0.5}_{0.25 \leq A_1 < 0.5}~\underbrace{\cdots}_{l_k \leq A_k < u_k} }^{\text{filter featurization~} \filterfeaturization}$ \\
	\end{tabular}
	\vspace{-12pt}
	\caption{An example of query featurization.}
	\label{fig:query_featurization_example}
\end{figure}
\end{small}

\vspace{-5pt}
\subsection{Discussions}
\label{subsec:featurization_discussion}

\subsubsection{Featurization properties.}
\label{subsubsec:featurization_discussion}

As we claimed, our way of featurizing database data and SQL queries has the following properties:

\noindent\underline{1) Efficiency.} Building the data featurizations of a static database requires looking over each relation once only. 
An insert/delete/update statement only influences one relation and modifying the relevant histograms takes $O(v)$ time where $v$ is the number of the records involved. 
The time complexity of featurizing a SQL query is $O(|\boldsymbol{J}| + |\boldsymbol{F}|)$, which is small and can often be ignored.

\noindent\underline{2) Low space overhead.} The DB states only contain $T \cdot \dx$ float numbers. In practice, we usually set $\dx$ smaller than 100. The dimensionality of a query featurization is $2m \cdot \Delta + 2T$. Usually, the value of $m$ is small. In our experiments, $m$ is smaller than 10. Our way of re-generating join predicates based on equivalence classes will usually reduce $\Delta$, the number of possible joins. Also, the joins tends to be performed on attributes with primary keys. According to our observations, $\Delta$ is usually less than $N^2$ and featurizing a query on a 8-relation database requires less than 1000 float numbers.

\noindent\underline{3) Stability.} The dimensions of data and query featurizations are fixed, no matter how the database or queries change. This ensures the featurizations can be easily processed with learned models.

\vspace{-6pt}
\subsubsection{Why our featurizations work on dynamic workloads?}
\label{subsubsec:featurization_for_dynamic_workload}

It is noteworthy that no matter how the underlying data changes, the featurization of a given query will not change. In contrast, the DB states will vary with the change of the data in the database. It is a compression of the whole database and able to catch the distribution characteristics of each attribute.
Our model takes both featurizations as input and is able to `convolute' the query featurizations with the DB states. Thus, when the data changes, it can react properly and give different predictions for the same query with different DB states. 
\if\includeAppendix1
The experiments in Section~\ref{sec:experiments} and Section~\labelAdditionalExp in Appendix show that our model outperforms other state-of-the-art methods on both static and dynamic workloads.
\else
The experiments in Section~\ref{sec:experiments} and Section~\labelAdditionalExp in the extended version~\cite{extended_url} show that our model outperforms other state-of-the-art methods on both static and dynamic workloads.
\fi

\vspace{-2pt}
\section{Design of~\modelName}
\label{sec:approach}
\vspace{-1pt}

Given the DB states $\dbstates$ and a SQL query featurization $\sqlq$, our \mname can reasonably discover the implicit relations between them that are required for cardinality estimation. The mystery behind lies in the attention mechanisms~\cite{DBLP:conf/nips/VaswaniSPUJGKP17} twice used in our~\modelName. In this section, we will review the motivation of~\modelName's design, introduce the details of~\modelName~including its two modules processing $\dbstates$ and $\sqlq$, respectively, and the training process of~\modelName.

 \vspace{-3pt}
\subsection{Motivations and~\modelName~Overview}
\label{subsec:motivations}
 \vspace{-1pt}

An attention function maps a query and a set of key-value pairs to an output, where the query, keys, values, and output are all vectors~\cite{DBLP:conf/nips/VaswaniSPUJGKP17}. 
Here, the concepts of key-value pairs and queries can be analogous to retrieval systems. 
Take a user's search behavior on an e-commerce platform like Amazon as an analogy. When the search engine receives a query (the text in the search bar),  it maps the query against a set of keys (item names, tags and descriptions, \emph{etc.}) associated with values (candidate items) in the database and outputs the best matched items.
The output is a weighted sum of the values, where each weight is computed by a compatibility function that measures the relevance between the query and keys.

\begin{figure}[htb]
	\centering
	\vspace{-2pt}
	\if\includeAppendix{1}
	\includegraphics[width=0.49\textwidth, trim={8mm 5mm 5mm 4mm},clip]{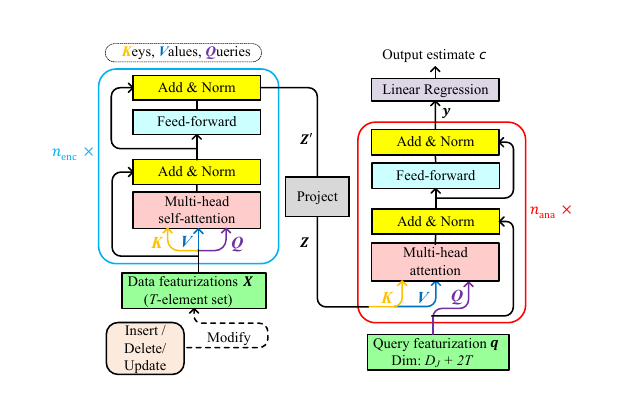}
	\else
	\includegraphics[width=0.49\textwidth, trim={8mm 5mm 5mm 4mm},clip]{model.eps}
	\fi
	\vspace{-19pt}
	\caption{~\modelName~structure.}
	\label{fig:model_structure}
	\vspace{-1pt}
\end{figure}

The idea behind the attention mechanism is to encode the input key-value pairs set, and utilize the most relevant parts of the keys, associated with values, with the query in a flexible manner. 
Through a weighted combination of all encoded input vectors, this mechanism `answers' the query with the most relevant vectors getting the highest weights. 
This idea perfectly fits in our research problem as a SQL query's featurization $\sqlq$ is a natural query vector. 
Besides, the DB states $\dbstates$ can be seen as the item information in the above example and be used as the keys and values in the attention functions. Thus, we extend this idea and design~\modelName, which takes full advantage of the attention mechanism to accurately estimate cardinalities of SQL queries on dynamic workloads. Fig.~\ref{fig:model_structure} illustrates the structure of~\modelName.

Our attention-based model~\modelName~is designed like a soft lookup table. It fetches information from a better representative of the DB states $\dbstates$ using features from the transformation of $\sqlq$ as indices. 
It is composed of two modules. The left \textbf{data-encoder} module maps $\dbstates = \{\adbstate_1, \cdots, \adbstate_T\}$ to another set of representations $\encout = \{\anencout_1, \cdots, \anencout_T\}$. 
It learns the implicit joint distribution information among all attributes by computing the relevance between any pairs of DB states through a stack of self-attention layers. This distribution information is embodied in the set $\encout$.
Given $\encout$, the right \textbf{query-analyzer} module adopts another stack of attention layers to measure the relevance between the query featurization $\sqlq$ and $\encout$, and generates an `answer' vector $\answervec$. 
Finally, vector $\answervec$ is fed into a linear regression layer and $\sqlq$'s cardinality given DB states $\dbstates$ is estimated and returned. 
 
 \begin{table*}[htb]
 	\vspace{-18pt}
 	\smaller
 	\centering
 	\caption{The overall picture of~\modelName's two modules, given DB states $\dbstates$ and SQL query featurization $\sqlq$}\label{tab:attention_elements}
 	\vspace{-10pt}
 	\setlength\tabcolsep{3pt}
 	\begin{tabular}{ l l l V{3} l l l} \hline
 		\multicolumn{3}{c V{3}}{\textsf{Data-encoder (Self-attention)}} & \multicolumn{3}{c} {\textsf{Query-analyzer (Data-query cross attention)}} \\ 
 		\textbf{\emph{Input source}} & \textbf{\emph{What to learn?}}  & \textbf{\emph{Output}} & \textbf{\emph{Input source}} & \textbf{\emph{What to learn?}}  & \textbf{\emph{Output}} \\ \hline 
 		\tabincell{l}{Keys: $\dbstates$\\Values: $\dbstates$\\Queries: $\dbstates$} 
 		& \tabincell{l}{Relevance among the DB states $\dbstates$,\\and thus the joint distribution inf-\\ormation of multiple attributes.} 
 		& \tabincell{l}{Another vector set $\encout$ cov-\\ering joint distribution\\ information of attributes}
 		& \tabincell{l}{Keys: $\encout$\\Values: $\encout$\\Queries: $\sqlq$}
 		& \tabincell{l}{Relevance between $\sqlq$ with the ele-\\ments in $\encout$, showing which parts\\of data are more important.} 
 		& \tabincell{l}{Final `answering' vector $\answervec$ wh-\\ich will be directly turned into\\the cardinality estimate $\card$.}\\\hline
 	\end{tabular}
 	\vspace{-15pt}
 \end{table*}

\vspace{-6pt}
\subsection{Attentions in~\modelName} 
\label{subsec:attention}
 \vspace{-2pt}

\noindent\textbf{\underline{Attention background.}} Before showing the data-encoder and query-analyzer modules in~\modelName, it is worth briefly introducing the attention mechanism.
In neural networks, attention is a technique that is meant to mimic cognitive attention. Its motivation behind is that the network should devote more focus to the important parts of the data instead of treating all data equally. It uses an attention function to discover which parts of the data should be emphasized. 
The function maps a query and a set of key-value pairs to an output, which is a weighted sum of the values.
Usually, the function computes the similarity (relevance) between each pair of query and key with some metric, and uses it to produce the weight assigned to the corresponding value. 

Our~\modelName~uses the `Scaled Dot-Product Attention'~\cite{DBLP:conf/nips/VaswaniSPUJGKP17}, namely $\mathsf{Attn}$, as the attention functions in the attention layers from both the data-encoder and query-analyzer modules. The input keys and values are packed together into matrices $\boldsymbol{K}$ and $\boldsymbol{V}$, of dimension $S \times d_k$ and $S \times d_v$, respectively, where $S$ denotes the size of the original key-value pair set.\footnote{In the rest of the paper, we regard a set of vectors as a matrix. In other words, we do not distinguish a set of key/value/query vectors with a key/value/query matrix.} The $d_k$-dim query vector is converted to a $1\times d_k$ matrix $\boldsymbol{Q}$. The function $\mathsf{Attn}$ computes the dot products of the given query $\boldsymbol{Q}$ with all keys $\boldsymbol{K}$, divides each by $\sqrt{d_k}$, and applies a softmax function to obtain the weights on the values $\boldsymbol{V}$:

\vspace{-11pt}
\begin{small}
\begin{flalign}\nonumber
\mathsf{Attn}(\boldsymbol{Q, K, V}) &= \text{softmax}\big(\frac{\boldsymbol{Q} \cdot \boldsymbol{K}^{T}}{\sqrt{d_k}}\big)\boldsymbol{V}
\end{flalign}
\end{small}
\vspace{-7pt}

The dot product above is used as the similarity metric. To prevent the dot products from growing too large such that the gradients of the softmax function `vanishing', we divide the dot product by the factor$\sqrt{d_k}$.
In practice, the function $\mathsf{Attn}$ is computed with batches of queries simultaneously to improve the efficiency.
Before feeding the input to $\mathsf{Attn}$, we adopt a multi-head projection mechanism~\cite{DBLP:conf/nips/VaswaniSPUJGKP17} to first project the queries, keys and values multiple times with $h$ different linear projections. This operation is able to enhance the representation ability of~\modelName. 
\if\includeAppendix1
Due to space limit, more details are given in Section~\labelMultihead in Appendix.
\else
Due to space limit, more details are given in Section~\labelMultihead in the extended version~\cite{extended_url}.
\fi

\noindent\textbf{\underline{Attentions in~\modelName's two modules.}}The multi-head attention layer appears twice in \modelName. The first is a self-attention layer in the data-encoder module. It is fed with the same query, key and value matrix $\dbstates$ of dimension $T$$\times$$\dx$, and outputs a matrix $\boldsymbol{Z'} \in \mathbb{R}^{T \times \dx}$ that will be later projected to another matrix $\encout \in \mathbb{R}^{T \times \dq}$. Then, $\encout$ is used as the key and value matrix of the other attention layer in the query-analyzer module, where the input query set contains only one SQL query featurization vector.
It is noteworthy that the "queries" set in the self-attention layers does not come from the SQL query. 
In addition, we do not need to order the vectors in the DB states. Thus, unlike transformer-like models to process sequences~\cite{DBLP:conf/naacl/DevlinCLT19,DBLP:conf/nips/BrownMRSKDNSSAA20}, \modelName~does not need the positional encodings~\cite{DBLP:conf/nips/VaswaniSPUJGKP17} or positional embeddings~\cite{DBLP:conf/icml/GehringAGYD17} of the attention layers' inputs.

Table~\ref{tab:attention_elements} states the input sources and output of the two modules' attention layers, and what each module is able to learn. The rationale behind them is given in the following two subsections.

\vspace{-4pt}
\subsection{Data-encoder} 
\label{subsec:data_encoder}
\vspace{-2pt}

\noindent Suppose a random variable $V$'s value is taken randomly with an equal possibility from the possible values of an attribute. As a result, the corresponding histogram can be regarded as the rough distribution function of $V$. Thus, the DB states consists of a series of `marginal' distributions describing single attributes. However, SQL queries tend to cover join predicates that need to know the joint distribution information among multiple attributes. Usually, the distributions of different attributes are not independent. It is difficult and impractical to directly derive the joint distribution from the marginal distributions. To address this issue, we design the data-encoder module that makes use of attentions, to establish a bridge between the marginal distributions and the joint distribution.

The data-encoder takes as input the DB states $\dbstates$ and feeds it into a $\boldsymbol{\nenc}$-stacked layers. 
Each layer is identical and has two sub-layers. The first is the multi-head self-attention sub-layer $\mathsf{N_{sa}}$ which takes the same inputs of keys, values and queries---three matrices equal to $\dbstates$ or the output of the last layer, and outputs another $T$-element set $\boldsymbol{\Tilde{Z}}_i$. On top of $\mathsf{N_{sa}}$, the feed-forward sub-layer $\mathsf{FF}$ uses stacked fully connected networks and nonlinear activation functions, \emph{e.g.,} ReLU, to map $\boldsymbol{\Tilde{Z}}_i$ into the data representation set $\encout'_i$. Besides, to address the degradation problem and ease the model training, we employ a residual connection~\cite{DBLP:conf/cvpr/HeZRS16} around each sub-layer, followed by layer normalization~\cite{DBLP:journals/corr/BaKH16}, following the settings in~\cite{DBLP:conf/nips/VaswaniSPUJGKP17}. Thus, the output of each sub-layer is LayerNorm($\boldsymbol{V} + \text{sub-layer}(\boldsymbol{V}))$. The sub-layer is either $\mathsf{N_{sa}}$ or $\mathsf{FF}$. The analytic expression of the output representations $\primeencout$ is described as follows.

\vspace{-13pt}
\begin{small}
\begin{flalign} \nonumber
\primeencout &= \encout_{\nenc}', \text{~where~~~} \encout_{0}' = \dbstates, \text{~and~} \forall 1 \leq i \leq \nenc, \\ \nonumber
\boldsymbol{\Tilde{Z}_{i}} &= \mathsf{LayerNorm}\big(\encout'_{i-1} + \mathsf{N_{sa}}(\overbrace{\encout_{i-1}'}^{\text{keys}}, \overbrace{\encout_{i-1}'}^{\text{values}}, \overbrace{\encout_{i-1}'}^{\text{queries}})\big), \\ \nonumber
\encout'_i &= \mathsf{LayerNorm}(\boldsymbol{\Tilde{Z}}_i + \mathsf{FF}(\boldsymbol{\Tilde{Z}}_i)),\text{with~} \mathsf{FF}(\boldsymbol{V}) = \text{ReLU}(\boldsymbol{WV} + \boldsymbol{b})
\end{flalign}
\end{small}
\vspace{-13pt}

Finally, the output matrix $\primeencout$ will be linearly projected to the representation matrix $\encout \in \mathbb{R}^{T \times \dq}$. This projection is to align the dimensionality of $\encout$ with that of the query featurization $\sqlq$.

The keys, values and queries in the attention layers are the same set. They are either the DB states $\dbstates$ or the output of the previous stacked layer. 
This setting makes each element in the output set of a layer attend to all outputs in the previous layer and thus attend to all DB states. 
More importantly, the self-attention layer quantitatively `calculates' the relevance between a pair of elements from any two histograms. It is noteworthy that each element of a histogram describes the local distribution of an attribute.
Therefore, the data-encoder module is able to more effectively discover the implicit connections between any pair of DB states, and thus exhibits behavior in relation to the joint distribution of multiple attributes in the output set. 
Compared to other neural network architectures like multilayer perceptron (MLP)~\cite{DBLP:books/lib/HastieTF09}, self-attention could yield more interpretability and have higher representative abilities.
Through self-attention layers, we create links among all DB states, or equivalently, all attributes in the database.  
After fine-

\noindent tuning the parameters of the self-attention and feed-forward layers, the relationship information helpful to the cardinality estimation task is implicitly covered and encoded into the output $\encout$ of the data-encoder module. 
This information will be processed and utilized in the query-analyzer module.

\vspace{-4pt}
\subsection{Query-analyzer} 
\label{subsubsec:query_analyzer}
\vspace{-1pt}

The query-analyzer module attempts to discover and measure the relevance, through data-query cross attention layers, between the SQL query and each element of $\encout$, the output of the data-encoder module covering joint distribution information among attributes.

This module is also a stacked structure composed of $\boldsymbol{\nana}$ identical layers. 
Similar to the data-encoder module, each layer here is composed of a multi-head attention sub-layer and a fully connected sub-layer. Also, residual connections are employed around each sub-layer, followed by layer normalization. 
Unlike the data-encoder module, the input sets of key-value pairs and queries to the `data-query' attention sub-layer here are not from the same place. 
We use $\encout$, the output of the data-encoder module, as both the input key and value matrices, while the query set here comes from either the query featurization or the output of the previous layer. 
It is noteworthy that the input query set and the output set of each attention sub-layer have only one $\dq$-dim element.

The data-query attention sub-layer establishes a bridge between the queries and the data. It individualize each SQL query featurization and presents different `answers' by enhancing the influences of some parts of the input key-value pair set $\encout$, while diminishing other parts. Learning which part of the data is more important depends on the relations between queries and keys, and this is measured with the attention functions. Suppose a SQL query contains join predicate $R_1.A^{1}_1 = R_2.A^{2}_2$ and filter predicate $R_1.A^{1}_2 > 1$, and the attributes $R_1.A^{1}_1$, $R_2.A^{2}_2$ and $R_1.A^{1}_2$ are numbered $i$, $j$ and $k$, respectively. The data-query attention sub-layer will pay more attention to the part of the vectors in the set $\encout$ that are relevant to $\adbstate_i$, $\adbstate_j$ and $\adbstate_k$. The effect of particular attention can be realized through suitable parameters of different layers in both modules. 

After accessing $\answervec$, the output of the final query-analyzer layer, we use a simple linear regression layer $\mathsf{LR}$ to calculate a scalar value as the cardinality estimate $\card$. The process of accessing $\card$ with the input of $\encout$ and $\sqlq$ is described as follows.

\vspace{-10pt}
\begin{small}
\begin{flalign} \label{Eq:cardest}
\card &= \mathsf{LR}(\answervec), \text{~where~~~} \answervec = \answervec_{\nana}, \answervec_0 = \sqlq, \text{~and~} \forall 1 \leq i \leq \nana, \\ \nonumber
\answervec'_{i} &= \mathsf{LayerNorm}\big(\answervec_{i-1} + \mathsf{N_{dq}}(\overbrace{\encout}^{\text{keys}}, \overbrace{\encout}^{\text{values}}, \overbrace{\answervec_{i-1}}^{\text{queries}})\big), \\ \nonumber
\answervec_{i} &= \mathsf{LayerNorm}(\answervec'_{i} + \mathsf{FF}(\answervec'_{i})) 
\end{flalign}
\end{small}

\if\includeAppendix1
\vspace{-13pt}
\else
\vspace{-13pt}
\fi
\subsection{Training of~\modelName}
\label{subsec:training}
\vspace{-1pt}

Fine-tuning the parameters of~\modelName~requires a training dataset of which each element is a 3-tuple $(\sqlq_i, \dbstates_i, \card'_i)$, where $\sqlq_i$ is the query featurization of a SQL query $\sqlq_i$ and $\dbstates_i$ is the associated DB states of the dynamic database. By executing $\sqlq_i$ on the database of which the DB states is $\dbstates_i$, we will get the true cardinality $\card'_i$, which will be used as the label. In practice, we will take the logarithm of $\card'_i$ to make the range of the labels not too large. 
Collecting the training dataset is not difficult. Usually, we only need to collect the feedback of executed queries on a dynamic database. 
Then, we will get the training dataset with three lists $\boldsymbol{\mathcal{X}} = [\dbstates_i]$, $\boldsymbol{Q} =  [\sqlq_i]$ and $\boldsymbol{\card'} = [\card'_i]$. They will be split into batches to train our~\modelName. 

We use the mean-weighted-squared-error function $\mathsf{MWSE}$ taking input of the batch card predictions $\boldsymbol{\card_b}$ and the true cards $\boldsymbol{\card'_b}$ as well as their weights $\boldsymbol{w_b}$ with batch size $B$ as the loss function, \emph{i.e.,} $\mathsf{MWSE}(\boldsymbol{\card_b}, \boldsymbol{\card'_b}) = \frac{1}{B}\sum_{i=1}^{B}w_{i}(\card_i-\card'_i)^{2}$. The parameters of the linear regression layer, the attention and feed-forward sub-layers in both modules are trained by gradient descent with batches in an end-to-end fashion. 
Here, the value $w_{i}$ is proportional to that of $\log \card_i$. In particular, $w_i = \frac{\log \card_i}{\sum_{j} \log \card_j}$. We use the weight $w_i$ in the loss function because it is usually beneficial to emphasize the queries with larger true cardinalities as their execution times tend to be longer.
The procedure of training~\modelName~using the these lists $\boldsymbol{\mathcal{X}}$, $\boldsymbol{Q}$ and $\boldsymbol{\card'}$ is detailed in Algorithm~\ref{alg:training}.

\begin{algorithm}
	\smaller
	\caption{\textproc{Train-\modelName}}
	\label{alg:training}

	\SetKwFunction{tepoch}{Train-Epoch}
	\SetKwProg{Fn}{Function}{:}{}
	\SetKwInOut{KwIn}{Input}
	\SetKwInOut{KwOut}{Output}
	\SetArgSty{textnormal}
	
	\KwIn{~Three data lists: $\boldsymbol{\mathcal{X}} = [\dbstates_i]$, $\boldsymbol{Q} =  [\boldsymbol{q_i}]$, $\boldsymbol{\card'} = [\card'_i]$ \\
				~The maximum number of training epochs $M_e$	
}
	\KwOut{Well-trained~\modelName} 
	
	Compute the weight vector $\boldsymbol{w} = [w_i]$ according to $\boldsymbol{\card'}$. \\
	\For{$i \leftarrow 0$ \KwTo $M_e$}{
		$\tepoch(\boldsymbol{\mathcal{X}}, \boldsymbol{Q}, \boldsymbol{\card'}, \boldsymbol{w})$ \\
		\If{$\text{\modelName~performs better on validation set}$} {
			Assign current parameters to~\modelName.
		}
	}
	\KwRet{\modelName~whose parameters are fine-tuned.}
 
	\Fn{\tepoch{$\boldsymbol{\mathcal{X}}, \boldsymbol{Q, \card', w}$}}{
		Shuffle and split $\boldsymbol{\mathcal{X}}, \boldsymbol{Q, c, w}'$, generate $n_B$ equal-size batches $[(\boldsymbol{\mathcal{X}}_b, \boldsymbol{Q}_b, \boldsymbol{\card'}_b), \boldsymbol{w}_b]$, $b = 1,\cdots, n_B$. \\
		\For{$b \gets 0$ \KwTo $n_B$}{
			Compute the vector $\boldsymbol{\card}_b$ with Eq (\ref{Eq:cardest}).\\
			Compute the $\mathsf{MWSE}$ loss given $\boldsymbol{\card}_b$, $\boldsymbol{\card'}_b$ and $\boldsymbol{w}_b$, and apply gradient descent to tune model parameters.
		}
	}
\end{algorithm}
In practice, a training dataset is split into two parts: the first part is the three data lists used to fit the parameters in the learning process (lines~7-10); the second part is used as the validation set to choose best parameters for~\modelName~and avoid overfitting (lines~4-5). Also, we will apply the early stopping strategy~\cite{DBLP:series/lncs/Prechelt12}, to stop the training process when the error on the validation set grows.

\vspace{-4pt}
\section{Analysis of~\modelName}
\label{sec:analysis}
\vspace{-2pt}

In this section, we discuss why~\modelName~is suitable for dynamic workloads and further analyze its properties.

\vspace{-6pt}
\subsection{\modelName~on Dynamic Workloads}
\label{subsec:model_on_dynamic_workloads}
\vspace{-1pt}

It is critical for databases to process dynamic workloads, \emph{i.e.,} mixture of queries and data manipulation operations. This requires a DBMS' query optimizer to be able to accurately predict the cardinalities of the (sub-)queries performed on a continuously changing database. To reduce the related overhead, the DBMS regularly maintains its

\noindent cardinality estimators instead of modifying it immediately after each update. Between two model-maintaining time points, the cardinality estimator model is usually fixed but still expected to make accurate estimates.

The traditional histogram-based cardinality estimators~\cite{DBLP:conf/sigmod/SelingerACLP79,DBLP:conf/vldb/PoosalaI97,DBLP:conf/sigmod/GunopulosKTD00,DBLP:conf/sigmod/DeshpandeGR01} currently used in systems like PostgreSQL~\cite{postgresql} make simple and often unreasonable assumptions, like the mutual independence of attributes. Their estimates are inaccurate, especially when the data is dynamic. This renders it difficult for the optimizer to choose good query plans. Existing learned cardinality estimation methods either do not support dynamic workloads~\cite{DBLP:conf/sigmod/ZhaoYHLZ22} or need to completely re-build the model at model-maintaining time points~\cite{DBLP:journals/pvldb/ZhuWHZPQZC21,DBLP:journals/pvldb/HilprechtSKMKB20,DBLP:journals/pvldb/YangKLLDCS20}. This process is often too time-consuming to be feasible.



Compared to existing methods~\cite{DBLP:journals/pvldb/HilprechtSKMKB20,DBLP:journals/pvldb/ZhuWHZPQZC21,DBLP:journals/pvldb/YangKLLDCS20}, the training of \modelName is efficient. Also, it is easily maintained: it can be incrementally updated at model-maintaining time points. 
Moreover, its architecture design is reasonable and suitable for processing dynamic workloads. 
It is noteworthy that the SQL queries and underlying data are decoupled in~\modelName~such that ~\modelName~can learn something whenever the DB states $\dbstates$ or the query featurization $\sqlq$ in the training dataset changes. Thus,~\modelName~is able to avoid overfitting certain datasets. Instead, it applies the attention mechanisms on its two modules to learn the properties of the data distribution which `generates' the underlying data, and how these properties influence the cardinality of a SQL query.
It is able to reasonably approximate  $\card(\sqlq, \mathsf{DB})$, the true but implicit mapping among the queries, dynamic data and cardinalities. When either the query featurization or DB states, or both of them, get changed, ~\modelName~can still output accurate results.
Our experimental results indicate that~\modelName~achieves good performance even if there is distribution discrepancy between the training and testing data.

\noindent\textbf{\underline{A straightforward baseline without attentions.}}
It is mentioned in Section~\ref{subsec:featurization_discussion} that the DB states will be dynamically modified when data is inserted to/deleted from the database. This renders it possible to use a stable well-trained model to predict the cardinalities of queries on a dynamic workload. 
As the dimensions of DB states and query featurizations are fixed, we can also adopt a straightforward method without attentions to process them. For example, for each pair of data featurizations $\dbstates_i$ and query featurization $\sqlq_i$, we flatten the matrix $\dbstates_i$ into a vector and concatenate the vector with $\sqlq_i$ to generate another vector $\boldsymbol{v}_i$. Subsequently, $\boldsymbol{v}_i$ and the associated layer are fed into a common supervised neural network like multilayer perceptron (MLP)~\cite{DBLP:books/lib/HastieTF09}. 
However, a straightforward neural network like MLP is usually not powerful enough to discover the implicit relations between SQL queries and the underlying data.
Its performance heavily relies on the similarity between the distributions of training and testing datasets. 
When the workload is static, \emph{i.e.,} the database data never changes, straightforward methods perform well.
However, the cardinality estimator model needs to use the current training data it observes to make predictions for `future' data. The distribution of future DB states may be highly different from that of the current data. 
In contrast, the application of the attention mechanisms in our~\modelName~make it possible to make accurate estimates even when the distributions of the underlying data change.
Experimental results in Section~\ref{sec:experiments} show the great advantages of~\modelName~over MLP on processing dynamic workloads.

\vspace{-10pt}
\subsection{Overhead of \modelName and Its Extension}
\label{subsec:approach_discussions}
\vspace{-2pt}

Several other good properties make~\modelName~more practical and attractive as a cardinality estimation method in modern DBMS's query optimizers.
First, the storage and training overhead of \modelName is affordable. On the one hand, the sizes of~\modelName~on the three datasets in our experiments are both smaller than 23 MB. On the other hand, training \modelName from scratch requires less than 12 minutes. Also, the estimation latency of \modelName is less than 11 ms. These overheads make \modelName able to process real world workloads.

Second, different from FLAT~\cite{DBLP:journals/pvldb/ZhuWHZPQZC21} and DeepDB~\cite{DBLP:journals/pvldb/HilprechtSKMKB20}, ~\modelName~directly estimates the cardinality instead of the selectivity of a SQL query. To get the cardinality, the selectivity estimation methods usually needs to estimate the size of the corresponding join table with sampling-based techniques~\cite{DBLP:conf/sigmod/ZhaoC0HY18}. However, when multiple relations are involved, the statistical variance often becomes large and results in highly inaccurate estimates.

Last but more importantly,~\modelName~establishes a general framework that is not limited to cardinality estimation tasks. 
By replacing $\mathsf{COUNT(*)}$ with with other aggregation functions, the special aggregate analytic queries of Format~(\ref{Eq:query_format}) can be further transformed to more general ones.
As a matter of fact,~\modelName~can be easily extended to approximate the results of more general aggregate analytic queries. 
When processing queries with other kinds of aggregate functions, we only need to make slight modifications to the data and query featurizations, including featurizing the extra aggregate functions and optional $\mathsf{GROUP~BY}$ clauses in the query featurizations, and incorporating more data description information specific to the aggregate queries. We left the support of the general aggregate analytic queries for future work.
%
\vspace{-2pt}
\section{Experimental Studies}
\label{sec:experiments}
\vspace{-1pt}

This section reports the experiments that compare~\modelName~with selected alternatives. 
All methods are implemented in Python 3.9 and evaluated on a Linux server with a 96-core Xeon(R) Platinum 8163 CPU and 376GB memory.
The implementation of~\modelName~is open~\cite{codes_url}.
\if\includeAppendix1
Due to space limit, we present additional experimental results and analyses in Section~\labelAdditionalExp in Appendix.
\else
Due to space limit, we present additional experimental results and analyses in Section~\labelAdditionalExp in the extended version~\cite{extended_url}.
\fi

\vspace{-4pt}
\subsection{Experimental Settings}
\label{subsec:exp_settings}
\vspace{-2pt}

\noindent\textbf{Datasets.} We use three real datasets to evaluate all models. 
\vspace{-2pt}
\begin{itemize}[leftmargin=*]
	\item \textbf{STATS} contains 8 relations (\emph{users, posts, postLinks, postHistory, comments, votes, badges, tags}) with 43 attributes~\cite{STATS_url}. There are 1,029,842 records in total. An existing open workload with 146 queries are marked as `testing queries'. They are associated with 2,603 sub-queries. We created another 1000 different queries with sub-queries, which are used as the training and validation data.
	\item \textbf{Job-light} is a subset of the IMDB dataset~\cite{imdb_url}. It contains 6 relations (\emph{cast\_info, movie\_info, movie\_companies, movie\_keyword, movie\_info\_idx, title}) with 14 attributes. There are 62,118,470 records in total. The testing query set contains 208 queries, associated with 3,248 sub-queries. Similarly, another 2,000 queries as well as their sub-queries are created for training the models.
	\item \textbf{TPCH} (1 GB)~\cite{tpch_url} is a widely-used benchmark dataset of a suite of business oriented relations (\emph{customer, lineitem, nation, orders, part, partsupp, region and supplier}). 
	We remove the string type `comment' attribute from each relation as it is not supported by \modelName or other methods. We use the remaining 46 attributes.
	There are 8,661,245 records in total. We randomly create 123 testing queries and 1,554 testing sub-queries for evaluation. Another 2,000 queries and their sub-queries are used for training.
\end{itemize}
\vspace{-2pt}

\if\includeAppendix1
The join information among relations in these datasets are shown in Fig.~\labelFigJoins in Appendix. 
\else
The join information among relations in these datasets are shown in Fig.~\labelFigJoins in the extended version~\cite{extended_url}. 
\fi
All joins in the queries on the Job-light dataset are PK-FK joins, while the queries on the other datasets involve more complex many-to-many joins.

\noindent\textbf{Dynamic workloads.} For each dataset, we create three different dynamic workloads, each of which is a random mix of inserts, deletes, updates and query statements. These workloads are differ-

\noindent entiated according to the  ratios among the numbers of inserts, deletes and updates, and the distributions of the inserted records:
\vspace{-2pt}
\begin{itemize}[leftmargin=*]
	\item \textbf{\textsf{Insert-heavy}:} \#inserts : \#deletes : \#updates = $2 : 1: 1$.
	\item \textbf{\textsf{Update-heavy}:} \#inserts : \#deletes : \#updates = $1 : 1: 2$.
	\item \textbf{\textsf{Dist-shift}} is a special \textsf{Insert-heavy} workload but having skewed distribution of the inserted records.
	In particular, the inserted records are selected intentionally such that their first attributes' values are the first 30\% smallest among all data. 
\end{itemize}
\vspace{-1pt}

To generate a dynamic workload, we randomly select about $\frac{2}{3}$ of the records as the initial datasets to bulk load all the relations.
Subsequently, the insert and update statements in a dynamic workload will insert the remaining $\frac{1}{3}$ of the records to the database, while the delete and update statements will remove or change some records. For simplicity and clarity, we assume that each delete or update statement only influences one record.
These statements are equally split into two parts: the former `training' part and the latter `evaluation' part. 
To ensure that \modelName learns from enough features, we make three copies of the training queries and their sub-queries and randomly mix them with the training part.
The training part of the workload is used as the training dataset for building all cardinality estimation methods.
Note that the true cardinalities of the queries in the training part are also available to all methods.

The way of mixing testing queries with the data manipulation statements in the evaluation part are slightly different. For each testing query, we pack it up with its sub-queries together. Then, those packs are shuffled and randomly mixed with data manipulation statements in the evaluation part. Also, we'd like to know how cardinality estimators perform when a certain amount of data in the database changes. 
Thus, it is assumed that when each testing query is executed, the associate \textbf{changing rate $\rho$} of the underlying database data is larger than a pre-defined threshold.
We use $H$ to denote the set of all records in the database after executing all the statements in the training part.
Given any query $q_i$ in the evaluation part, suppose $B_i$ is the set of all records in the database when $q_i$ is executed, and the numbers of inserts, deletes and updates from the evaluation part so far are $\mathtt{I_i}$, $\mathtt{D_i}$ and $\mathtt{U_i}$, respectively. The changing rate $\rho_i$ is defined as $\rho(t) = \frac{|H \triangle B_i|}{|H|} = \frac{\mathtt{I_i} + \mathtt{D_i} + 2\mathtt{U_i}}{|H|}$, where $\triangle$ denotes the symmetric difference operation of two sets. When there is no ambiguity, the subscript $i$ is omitted.

Both training and testing queries are randomly generated. In particular, to generate a query, we run a series of Bernoulli tests with $p$$=$$0.5$ to determine which relations appear. Then, we enumerate all possible subsets of the join conditions among the selected relations such that they can be connected by these join conditions. One of the subsets is randomly selected as the join predicate set $\boldsymbol{J}$.
The attributes in the filter predicates $\boldsymbol{F}$ and their operators ($\le, \geq$, \etc) are determined with the analogous Bernoulli tests.
The right hand side values of the filter predicates are randomly sampled from the initial dataset.

\noindent\textbf{Competitors.} We include the following representative methods:
\vspace{-5pt}
\begin{itemize}[leftmargin=*]
	\item \textbf{PG} is the simplest 1D histogram based cardinality estimation method used in PostgreSQL~\cite{postgresql}.
	\vspace{-1pt}
	\item \textbf{Uni-Samp}~\cite{DBLP:conf/sigmod/0002SSK21} uniformly samples a set of tuples with a given probability $p$ for each relation. The cardinality estimate of a query $q$ equals ${N_q}/{p^m}$ where $N_q$ is the number of returned tuples by executing $q$ on the sample sets and $m$ is the number of involved relations in $q$. We tune the value of $q$ to make Uni-Samp's storage overhead or latency is comparable to that of the other methods. The values of the sampling ratio $p$ for the STATS, Job-light and TPCH datasets are set to 0.1, 0.06 and 0.04, respectively.
	\vspace{-1pt}
	\item \textbf{DeepDB}~\cite{DBLP:journals/pvldb/HilprechtSKMKB20} is based on a Sum-Product-Network (SPN)~\cite{DBLP:conf/uai/PoonD11}. It learns a pure data-driven model to capture the data's joint probability distribution. Following the same settings in~\cite{DBLP:journals/pvldb/HilprechtSKMKB20}, we set the RDC independence threshold to 0.3 and split each SPN node with at least 1\% of the input data.
	\vspace{-1pt}
	\item \textbf{FLAT}~\cite{DBLP:journals/pvldb/ZhuWHZPQZC21} improves SPN based on factorize-sum-split-product network (FSPN)~\cite{DBLP:journals/corr/abs-2011-09020}, a graphical model, to adaptively model the joint probability density function of attributes. It is data-driven.
	\item \textbf{NeuroCard}~\cite{DBLP:journals/pvldb/YangKLLDCS20} is a data-driven method, extending Naru~\cite{DBLP:journals/pvldb/YangLKWDCAHKS19} into the multi-table case, while Naru is a Deep Auto-Regression (DAR)~\cite{DBLP:conf/icml/GregorDMBW14} based cardinality estimation algorithm for a single table. The sampling size of the NeuroCard model is set to 8,000, following the settings in the paper.
	\item \textbf{FactorJoin}~\cite{DBLP:journals/pacmmod/WuNAKM23} combines classical join-histogram methods with learned single table cardinality estimates into a factor graph.
	\item \textbf{MLP}~\cite{10.5555/104279.104293} is the baseline neural network based method introduced in Section~\ref{subsec:model_on_dynamic_workloads}.
	\item \textbf{MSCN}~\cite{DBLP:conf/cidr/KipfKRLBK19} is a multi-set convolutional network which adopts the information from both queries and data.
	\item \textbf{NNGP}~\cite{DBLP:conf/sigmod/ZhaoYHLZ22} adopts the Neural Network Gaussian Process (NNGP) model~\cite{DBLP:conf/iclr/LeeBNSPS18} to learn from queries as well as their true cardinalities in a supervised manner.
\end{itemize}
\vspace{-3pt}
In addition, we also include the comparison with the results generated by true cardinalities. This \textbf{Optimal} method measures the best performance a method can achieve.

These competitors are chosen because they have better overall performance over other statistical and learned cardinality estimators. The comparisons are reported in benchmark and evaluation papers~\cite{DBLP:journals/pvldb/HanWWZYTZCQPQZL21,DBLP:journals/pvldb/SunZSLT21} and other existing works like~\cite{DBLP:journals/pvldb/ZhuWHZPQZC21,DBLP:journals/corr/abs-2012-14743,DBLP:journals/pvldb/YangKLLDCS20}.
Table~\ref{tab:model_properties} summarizes the properties of all methods. The better performance is indicated in bold.
\setlength{\textfloatsep}{5pt}
\setlength{\columnsep}{3pt}
\setlength{\intextsep}{3pt}
\begin{smaller}
	\begin{table}[htb]
		\centering
		\caption{Properties of different methods}
		\label{tab:model_properties}
			\vspace{-11pt}
		\setlength\tabcolsep{3pt}
            \begin{tabular}{l|c|c|c|c|c}
			\hline 
			\textbf{Method} &  \tabincell{c}{Data-\\driven} &  \tabincell{c}{Query-\\driven} & \tabincell{c}{Building\\time} & \tabincell{c}{Space\\overhead} & \tabincell{c}{Suitable for dy-\\namic workload}     \\ \hline
			PG~\cite{postgresql}& $\boldsymbol{\checkmark}$ & $\times$ &   \textbf{small}  & \textbf{low} & $\times$ \\ \hline
			Uni-Samp~\cite{DBLP:conf/sigmod/0002SSK21} & $\boldsymbol{\checkmark}$ & $\times$ & \textbf{small} & medium & $\times$ \\ \hline
			DeepDB~\cite{DBLP:journals/pvldb/HilprechtSKMKB20} & $\boldsymbol{\checkmark}$ & $\times$ & medium & medium & $\times$ \\ \hline
			FLAT~\cite{DBLP:journals/pvldb/ZhuWHZPQZC21} & $\boldsymbol{\checkmark}$ & $\times$ & large & high & $\times$  \\ \hline
			NeuroCard~\cite{DBLP:journals/pvldb/YangKLLDCS20} & $\boldsymbol{\checkmark}$ & $\times$ & medium & medium & $\times$ \\ \hline
			FactorJoin~\cite{DBLP:journals/pacmmod/WuNAKM23} & $\boldsymbol{\checkmark}$ & $\times$ & medium & \textbf{small} & $\times$ \\ \hline
			MLP~\cite{10.5555/104279.104293}  & $\times$ & $\boldsymbol{\checkmark}$ & \textbf{small}  & \textbf{low}  & $\times$ \\ \hline
			MSCN~\cite{DBLP:conf/cidr/KipfKRLBK19} & $\boldsymbol{\checkmark}$ & $\boldsymbol{\checkmark}$  & \textbf{small}& \textbf{small} & $\times$ \\ \hline
			NNGP~\cite{DBLP:conf/sigmod/ZhaoYHLZ22}  & $\times$ & $\boldsymbol{\checkmark}$ & \textbf{small}  & medium & $\times$ \\ \hline
			\modelName~(ours) & $\boldsymbol{\checkmark}$ & $\boldsymbol{\checkmark}$  & \textbf{small}  & medium & $\boldsymbol{\checkmark}$  \\ \hline
		\end{tabular}
		\vspace{-3pt}
	\end{table}
\end{smaller}

\begin{scriptsize}
\begin{table*}[htb]
		\vspace{-26pt}
		\centering
		\caption{Overall performance of cardinality estimation models on dynamic workloads}
		\label{tab:overall_results}
		\vspace{-11pt}
			\setlength\tabcolsep{2.5pt}
		\arrayrulecolor{black}
			\begin{tabular}{  L{0.6cm} V{2}  L{0.9cm} V{2}  C{0.68cm} | C{0.6cm}  C{0.6cm}  C{0.6cm}  C{0.6cm} | C{0.45cm}  C{0.45cm}  C{0.45cm}  C{0.6cm} | C{0.55cm} |  C{0.87cm} | C{0.65cm}  V{2}  C{0.68cm} | C{0.6cm}  C{0.6cm}  C{0.6cm}  C{0.6cm} | C{0.45cm}  C{0.45cm}  C{0.45cm}  C{0.6cm} }
				\specialrule{1pt}{0pt}{0pt}
				\multirow{3}{*} {Data} & \multirow{3}{*}{Model} & \multicolumn{12}{c V{3}}{\textsf{Insert-heavy}} & \multicolumn{9}{c}{\textsf{Update-heavy}}  \\ \hhline{*{2}{|~}*{21}{|-}|}
				& & \multirow{2}{*}{\tabincell{c}{E2E \\ Time(S)}} & \multicolumn{4}{c |}{Q-error} & \multicolumn{4}{c|}{P-error} & \multirow{2}{*}{\tabincell{c}{Size \\ (MB)}} & \multirow{2}{*}{\tabincell{c}{Building \\ Time(Min)} } & \multirow{2}{*}{\tabincell{c}{Latency \\ (ms)}} & \multirow{2}{*}{\tabincell{c}{E2E \\ Time(S)}} & \multicolumn{4}{c |}{Q-error} & \multicolumn{4}{c }{P-error}   \\ \hhline{*{3}{|~}*{4}{|-}|} \hhline{*{7}{|~}*{4}{|-}|} \hhline{*{15}{|~}*{4}{|-}|} \hhline{*{19}{|~}*{4}{|-}|}
				& & & 50\% & 90\% & 95\% & 99\% & 50\% & 90\% & 95\% & 99\% & & & & & 50\% & 90\% & 95\% & 99\% & 50\% & 90\% & 95\% & 99\% \\ \specialrule{1pt}{0pt}{0pt}
				\multirow{10}{*} {STATS} 
				& PG & \cellcolor{grey}$7,790$ & $190$ & $1.4$$\cdot$$10^5$ & $1.1$$\cdot$$10^6$ & $1.8$$\cdot$$10^7$ & $2.60$ & $25.44$ & $41.23$ & $243$ & - & - & - &\cellcolor{grey}$4,337$ & $524$ & $43,096$ & $3.7$$\cdot$$10^5$ & $1.1$$\cdot$$10^7$ & $1.78$ & $19.52$ & $29.01$ & $178$  \\  
				& Uni-Samp & \cellcolor{green2}$6,646$ & $1.35$ & $12.47$ & >$10^{10}$ & >$10^{10}$  & $2.71$ & $16.77$ & $21.65$ & $201$ & $3.34$ & $\boldsymbol{0.02}$ & $239$ & $6,002$ & $\boldsymbol{1.27}$ & $9.3$$\cdot$$10^9$ & >$10^{10}$ & >$10^{10}$ & $1.10$ & $12.86$ & $33.19$ & $153.11$\\
				& NeuroCard & >$30,000$ & $17.37$ & $1,388$ & $7,402$ & $3.0$$\cdot$$10^5$ & $2.38$ & $39.18$ & $824$ & $84,415$ & $90.76$ & $23.69$ & $32.42$ & $22,193$ & $18.49$ & $1,252$ & $6,784$ & $3.1$$\cdot$$10^5$ & $2.10$ & $31.75$ & $347$ & $11,730$ \\ 
				& FLAT & >$30,000$ & $12.77$ & $1,979$ & $12,897$ & $8.6$$\cdot$$10^5$ & $2.22$ & $70.08$ & $2,202$ & $8,738$ & $210.33$ & $53.77$ & $13.67$ & $24,319$ & $16.00$ & $2,376$ & $17,811$ & $9.4$$\cdot$$10^5$ & $2.20$ & $29.76$ & $724$ & $1.4$$\cdot$$10^5$ \\ 
				& FactorJoin & >$30,000$  & $22.62$ & $2,593$ & $31,936$ & $1.6$$\cdot$$10^6$ & $2.35$ & $66.12$ & $876$ & $8,384$ & $1.64$ & $0.42$ & $1.33$ & >$30,000$ & $23.76$ & $2,939$ & $36,198$ & $2.4$$\cdot$$10^6$ & $2.75$ & $115$ & $912$ & $4,409$  \\
				& MLP & >$30,000$ & $2.4$$\cdot$$10^6$ & >$10^{10}$ & >$10^{10}$ & >$10^{10}$ & $7.35$ & $491$ & $2,294$ & $6,887$  & $8.63$ & $3.11$ & $3.17$ & >$30,000$ & $2.3$$\cdot$$10^6$ & >$10^{10}$ & >$10^{10}$ & >$10^{10}$ & $6.66$ & $502$ & $3,025$ & $11,584$ \\
				& MSCN & $27,758$ & $20.09$ & $2,870$ & $17,037$ & $2.6$$\cdot$$10^5$ & $2.46$ & $87.23$ & $736$ & $8,738$ & $\boldsymbol{1.61}$ & $11.87$ & $\boldsymbol{1.02}$ & $25,766$ & $18.89$ & $2,382$ & $16,947$ & $1.8$$\cdot$$10^6$& $2.56$ & $64.64$ & $246$ & $46,721$  \\
				& NNGP & $12,883$ & $9.88$ & $827$ & $4,652$ & $2.6$$\cdot$$10^5$ & $1.41$ & $5.31$ & $27.33$ & $3,588$ & $32.44$ & $0.89$ & $32.66$ & $11,985$ & $8.15$ & $567$ & $3,639$ & $2.5$$\cdot$$10^5$ & $1.52$ & $5.62$ & $13.81$ & $2,753$\\ 
				& \modelName & \cellcolor{green4}$\boldsymbol{2,901}$ & $\boldsymbol{1.29}$ & $\boldsymbol{5.07}$ & $\boldsymbol{11.54}$ & $\boldsymbol{62.52}$ & $\boldsymbol{1.07}$ & $\boldsymbol{1.34}$ & $\boldsymbol{1.59}$ & $\boldsymbol{2.32}$ & $22.31$ & $9.25$ & $8.25$ & \cellcolor{green4}$\boldsymbol{2,402}$ & $1.42$ & $\boldsymbol{5.06}$ & $\boldsymbol{10.40}$ & $\boldsymbol{44.69}$ & $\boldsymbol{1.08}$ & $\boldsymbol{1.45}$ & $\boldsymbol{2.13}$ & $\boldsymbol{2.86}$ \\ 
				& Optimal& \cellcolor{aqua}$2,871$ & $1$ & $1$ & $1$ & $1$ & $1$ & $1$ & $1$ & $1$ & -  & - & - &\cellcolor{aqua}$2,349$ & $1$ & $1$ & $1$ & $1$ & $1$ & $1$ & $1$ & $1$  \\ \specialrule{1pt}{0pt}{0pt}
				\multirow{11}{*} {\tabincell{l}{Job-\\light}} 
				& PG & \cellcolor{grey} $6,128$ & $1.59$ & $7.14$ & $20.85$ & $223$ & $1.10$ & $1.32$ & $1.48$ & $2.89$ & - & - & - &\cellcolor{grey}$4,747$ & $1.67$ & $6.20$ & $14.09$ & $70.13$ & $1.10$ & $1.39$ & $1.64$ & $3.06$ \\  
				& Uni-Samp & $6,401$ & $1.23$ & $2.59$ & $6.53$ & >$10^{10}$ & $1.08$ & $1.85$ & $2.79$ & $146$ & $23.95$ & $\boldsymbol{0.52}$ & $163$ & $\cellcolor{green2}4,491$ & $1.33$ & $3.61$ & $7.50$ & >$10^{10}$ &$1.12$ & $1.71$ & $2.26$ & $3.67$ \\
				& DeepDB & $26,636$ & $11.31$ & $663$ & $6,328$ & $5.4$$\cdot$$10^5$ & $1.66$ & $13.71$ & $26.01$ & $49.57$ & $11.52$ & $11.93$ & $11.92$ & $23,190$ & $10.64$ & $628$ & $4,883$ & $8.8$$\cdot$$10^5$ & $1.50$ & $11.81$ & $27.71$ & $135$   \\
				& NeuroCard & $27,744$ & $14.73$ & $978$ & $6,766$ & $4.7$$\cdot$$10^5$ & $1.81$ & $12.40$ & $37.25$ & $144$ & $42.93$ & $9.11$ & $15.90$ & $16,553$ & $13.49$  & $1,000$ & $6,489$ & $1.5$$\cdot$$10^6$ & $1.60$ & $15.10$ & $25.94$ & $50.68$  \\ 
				& FLAT & $23,876$ & $11.22$ & $626$ & $5,786$ & $4.4$$\cdot$$10^5$ & $1.66$ & $13.98$ & $27.26$ & $89.67$& $11.10$ & $17.83$ & $3.96$ & $22,531$ & $10.13$ & $490$ & $4,221$ & $1.1$$\cdot$$10^6$ & $1.42$ & $12.87$ & $34.54$ & $195$ \\ 
				& FactorJoin & $16,665$ & $25.14$ & $3,456$ & $27,938$ & $2.5$$\cdot$$10^6$ & $1.52$ & $5.45$ & $10.23$ & $47.09$ & $4.66$ & $26.22$ & $1.33$ & $13,071$ & $27.31$ & $6,134$ & $36,155$ & $2.2$$\cdot$$10^6$ & $1.42$ & $4.83$ & $9.43$ & $22.75$ \\
				& MLP & \cellcolor{green2}$5,610$ & $4.30$ & $23.14$ & $59.37$ & $2,411$ & $1.13$ & $1.70$ & $2.10$ & $3.12$ & $6.19$ & $2.68$ & $2.28$ & $5,351$ & $5.59$ & $62.02$ & $138$ & $1,441$ & $1.17$ & $2.02$ & $2.88$ & $4.09$  \\
				& MSCN & $26,665$ & $24.42$ & $1,809$ & $9,217$ & $1.4$$\cdot$$10^5$ & $2.07$ & $20.98$ & $28.58$ & $106$ & $\boldsymbol{1.56}$ & $33.37$ & $\boldsymbol{0.77}$ & $29,118$ & $26.77$ & $1,448$ & $7,980$ & $1.5$$\cdot$$10^5$ & $2.59$ & $19.87$ & $61.18$ & $257$  \\
				& NNGP & $15,430$ & $4.5$$\cdot$$10^8$ & >$10^{10}$ & >$10^{10}$ & >$10^{10}$& $3.57$ & $8.39$ & $11.97$ & $24.47$ & $32.44$ & $0.83$ & $29.86$ & $4,959$ & $4.99$ & $59.86$ & $190$ & $50,036$ & $1.20$ & $2.05$ & $2.99$ & $4.96$ \\ 
				& \modelName & \cellcolor{green4}$\boldsymbol{5,168}$ & $\boldsymbol{1.14}$ & $\boldsymbol{2.08}$ & $\boldsymbol{3.85}$ & $\boldsymbol{14.82}$ & $\boldsymbol{1.07}$ & $\boldsymbol{1.24}$ & $\boldsymbol{1.32}$ & $\boldsymbol{1.77}$ & $22.25$ & $7.73$ & $7.28$ & \cellcolor{green4}$\boldsymbol{3,839}$ & $\boldsymbol{1.09}$ & $\boldsymbol{1.93}$ & $\boldsymbol{3.63}$ & $\boldsymbol{16.28}$ & $\boldsymbol{1.07}$ & $\boldsymbol{1.23}$ & $\boldsymbol{1.29}$ & $\boldsymbol{1.57}$  \\ 
				& Optimal& \cellcolor{aqua} $5,063$ & $1$ & $1$ & $1$ & $1$ & $1$ & $1$ & $1$ & $1$ & - & - & - &\cellcolor{aqua} $3,792$ & $1$ & $1$ & $1$ & $1$ & $1$ & $1$ & $1$ & $1$   \\ \specialrule{1pt}{0pt}{0pt}
				\multirow{10}{*} {TPCH} 
				& PG & \cellcolor{grey}$\boldsymbol{3,230}$ & $1.35$ & $4.88$ & $7.90$ & $16.16$ & $1.44$ & $\boldsymbol{1.57}$ & $\boldsymbol{1.59}$ & $\boldsymbol{1.62}$ & - & - & - &\cellcolor{grey}$2,385$ & $1.33$ & $5.01$ & $6.87$ & $14.64$ & $\boldsymbol{1.41}$ & $\boldsymbol{1.56}$ & $\boldsymbol{1.59}$ & $\boldsymbol{1.62}$ \\  
				& Uni-Samp & >$30,000$ & $1.22$ & $2.94$ & >$10^{10}$ & >$10^{10}$ & $1.81$ & $2.22$ & $45.60$ & $514$ & $23.29$ & $\boldsymbol{0.17}$ & $24.52$ & >$30,000$ & $1.30$ & $3.41$ & >$10^{10}$ & >$10^{10}$  & $1.75$ & $19.35$ & $50.34$ & $211$ \\
				& NeuroCard & $10,616$ & $43.43$ & $7,071$ & $27,735$ & $3.5$$\cdot$$10^5$ & $3.75$ & $10.91$ & $22.48$ & $106$ & $704.98$  & $34.38$ & $34.77$  & $6,562$ & $45.73$ & $5,896$ & $20,931$ & $1.2$$\cdot$$10^5$ & $3.16$ & $11.04$ & $16.49$ & $245$   \\ 
				& FLAT & $11,428$ & $46.04$ & $12,236$ & $70,383$ & $8.4$$\cdot$$10^7$ & $3.56$ & $10.91$ & $25.71$ & $1,182$ & $168.65$ & $41.42$ & $10.06$  & $7,864$ & $42.32$ & $7,328$ & $26,470$ & $3.7$$\cdot$$10^5$ & $3.56$ & $15.82$ & $21.70$ & $425$  \\ 
				& MLP & $7,261$ & $51,475$ & $1.4$$\cdot$$10^7$ & $3.5$$\cdot$$10^7$ & $4.7$$\cdot$$10^9$ & $4.71$ & $14.11$ & $22.74$ & $36.09$ & $8.63$ & $3.57$ & $3.62$ & $8,342$ & $79,317$ & $3.7$$\cdot$$10^7$ & $9.1$$\cdot$$10^7$ & >$10^{10}$ & $5.12$ & $17.63$ & $32.74$ & $63.39$   \\
				& MSCN & $7,718$ & $36.36$ & $6,709$ & $26,744$ & $4.4$$\cdot$$10^5$ & $3.58$ & $15.11$ & $25.71$ & $387$& $\boldsymbol{1.58}$ & $23.43$ & $\boldsymbol{0.79}$ & $8,353$ & $41.66$ & $5,457$ & $21,433$ & $1.7$$\cdot$$10^5$ & $3.80$ & $14.86$ & $25.69$ & $472$  \\
				& NNGP & $18,618$ & $1.3$$\cdot$$10^9$ & >$10^{10}$ & >$10^{10}$ & >$10^{10}$ & $5.56$ & $47.40$ & $59.18$ & $99.55$ & $32.44$ & $0.92$ & $41.76$ & $3,128$ & $25.88$ & $432$ & $986$ & $4,968$ & $1.54$ & $7.98$ & $10.90$ & $20.66$ \\ 
				& \modelName & $3,233$ & $\boldsymbol{1.15}$ & $\boldsymbol{2.29}$ & $\boldsymbol{3.37}$ & $\boldsymbol{8.86}$ & $\boldsymbol{1.43}$ & $\boldsymbol{1.57}$ & $1.60$ & $\boldsymbol{1.62}$ & $22.34$ & $11.62$ & $10.35$ & \cellcolor{green4}$\boldsymbol{2,380}$ & $\boldsymbol{1.25}$ & $\boldsymbol{2.11}$ & $\boldsymbol{2.95}$ & $\boldsymbol{10.87}$ & $\boldsymbol{1.41}$ & $\boldsymbol{1.56}$ & $\boldsymbol{1.59}$ & $\boldsymbol{1.62}$  \\ 
				& Optimal& \cellcolor{aqua}$3,227$ & $1$ & $1$ & $1$ & $1$ & $1$ & $1$ & $1$ & $1$ & -  & - & - &\cellcolor{aqua}$2,378$ & $1$ & $1$ & $1$ & $1$ & $1$ & $1$ & $1$ & $1$  \\ \specialrule{1pt}{0pt}{0pt}
			\end{tabular}
			\vspace{-16pt}
\end{table*} 
\end{scriptsize}

\noindent \textbf{Evaluation Metrics.} We use three metrics to evaluate all methods:
\vspace{-13pt}
\begin{itemize}[leftmargin=*]
\item \textbf{E2E time} is the total execution time of all queries in the evaluation part of a workload, in which we feed the query optimizer the estimated cardinalities of all relevant sub-queries. 
Those estimation results are acquired using cardinality estimation methods. It is the most important evaluation metric as it directly connects to the query optimizer and objectively shows if a cardinality estimation method could help improve the query performance of a DBMS.
This type of evaluation requires an improved benchmark over the existing work~\cite{DBLP:journals/pvldb/HanWWZYTZCQPQZL21}. This benchmark can integrate the estimation results by an external method to PostgreSQL.
\if\includeAppendix1
We show its details in Section~\labelBenchmark in Appendix.
\else
We show its details in Section~\labelBenchmark in the extended version~\cite{extended_url}.
\fi

\item \textbf{P-error}~\cite{DBLP:journals/pvldb/HanWWZYTZCQPQZL21} measures the gap between the optimal query plan and the generated plan based on the estimated cardinalities, without executing the given query. 
In particular, given a query $\sqlq$, a query plan $P$ and the set $\boldsymbol{\card}$ of estimated/true cardinalities of $\sqlq$'s all sub-queries, the DBMS will output an estimated cost $\mathcal{C}(P, \boldsymbol{\card})$ with a cost function $\mathcal{C}$. 
By feeding the query optimizer $\boldsymbol{\card}^T$, the set of true cardinalities of $\sqlq$'s all sub-queries, we can get the optimal plan $P^T$ for $\sqlq$.
Similarly, a cardinality estimation method $\mathsf{A}$ outputs for $\sqlq$ a set $\boldsymbol{\card}^E$ which results in another query plan $P^E$.
Then, $\text{P-error}(\mathsf{A}, \sqlq) = \frac{\mathcal{C}(P^E, \boldsymbol{\card}^T)}{\mathcal{C}(P^T, \boldsymbol{\card}^T)}$, where the denominator $\mathcal{C}(P^T, \boldsymbol{\card}^T)$ is the optimal execution cost, 
and the numerator $\mathcal{C}(P^E, \boldsymbol{\card}^T)$ is the cost by feeding the true cardinalities of sub-queries to the query plan generated by method $\mathsf{A}$. In other words, $\mathcal{C}(P^E, \boldsymbol{\card}^T)$ is the actual execution cost of $\sqlq$ if method $\mathsf{A}$ is adopted. 

\item \textbf{Q-error}~\cite{DBLP:journals/pvldb/MoerkotteNS09} measures the distance between the estimated cardinality $P$ and the true cardinality $T$ of a query. In particular, $\mathsf{Q}\text{-}\mathsf{error}(P,  T) = \max(\frac{T}{P}, \frac{P}{T})$.

\item \textbf{Storage overhead} is the memory size used by a method.

\item \textbf{Building time} indicates the offline training time of query-driven methods or construction time of data-driven methods.

\item \textbf{Estimation latency} is the average estimation time per sub-query used by a cardinality estimation method.
\end{itemize}
\vspace{-3pt}
As the optimizer only requires the cardinality estimates of the sub-queries, our evaluations also involve the testing sub-queries only. 

\noindent \textbf{Parameter Settings.}
We use a 40-bin-histogram for each attribute, \emph{i.e.}, $\dx = 40$. The values of $\nenc$ and $\nana$, \emph{i.e.}, the number of stacked attention layers in the data-encoder and query-analyzer, respectively, are both set to 4. To train our~\modelName, we use an Adam optimizer~\cite{DBLP:journals/corr/KingmaB14} with a learning rate of 0.01 and a batch size of 128.

\vspace{-6pt}
\subsection{Performance on Dynamic Workloads}
\label{subsec:exp_dynamic_workload}
\vspace{-3pt}

For each dynamic workload, we build all methods using the data in the training part to make estimates for the testing sub-queries in the evaluation part.
The featurizations of the training queries and sub-queries as well as the corresponding DB states and true cardinalities in the training part form the training data.
We use the training data to train the query-driven models including~\modelName, MLP, MSCN and NNGP. The data-driven models, namely DeepDB, NeuroCard, FLAT and FactorJoin, are built with the database data after all the statements in the training part of the workload are executed. This setting is compatible with real world scenarios as the cardinality estimation models are updated at regular intervals.
In our experiments, the associated changing rate $\rho$ of each testing query is larger than 20\%. In other words, when testing queries are executed, at least 20\% of the underlying data are changed, compared 

\noindent to the database when cardinality estimators are built.

The estimation results by different methods will be fed into our improved benchmark to compare the end-to-end query times (E2E times) of all methods on dynamic workloads. 
To investigate the performance gap between these methods and the optimal, we also feed the true cardinalities to the benchmark to get the optimal execution time. We do not compare with DeepDB on the STATS and TPCH datasets since it supports PK-FK joins only. 
The open implementation of Factorjoin is hard-coded for the STATS and Job-light datasets and does not support the TPCH dataset. Thus, Factorjoin is not tested on this dataset.
Besides the E2E time, we also record the Q-error and P-error distributions, building time, storage overhead and latency of all methods on different workloads. 
These results are shown in Table~\ref{tab:overall_results}. The storage overhead, building time and latency results on the \textbf{Update-heavy} workloads are similar to the counterparts on the \textsf{Insert-heavy} workloads and thus omitted.

\noindent\underline{\textbf{End-to-end evaluations.}} Referring to Table~\ref{tab:overall_results}, ~\modelName~has clear advantages on the E2E time over the other methods.
On the one hand, compared to PostgreSQL's built-in cardinality estimator, \modelName makes query execution up to 2.7$\times$ faster. 
Its E2E time is only slightly larger than that of PG on TPCH's \textsf{Insert-heavy} workload. However, there is almost no gap between the performance of PG and Optimal on this workload.
Compared to Optimal,~\modelName~only results in at most 2.2\% extra E2E time.
Considering the superiority of~\modelName~on these workloads, we confidently claim that~\modelName~performs much better than the built-in estimator in PostgreSQL and is able to greatly improve the query execution performance on dynamic workloads.
On the other hand, in addition to~\modelName, only MLP and Uni-Samp outperform PG on few workloads.
The E2E time of DeepDB, NeuroCard, FLAT, FactorJoin and NNGP is larger than that of PG in all cases. 
Compared to Optimal, these methods results in more than double of E2E time on most workloads. These results indicate that these existing methods cannot make satisfactory estimates for DBMS on the dynamic workloads.

The P-error comparisons demonstrate that~\modelName~results in ef-

\vspace{-1pt}
\noindent fective query executions from another angle. The P-errors of~\modelName on most queries are close to 1. In other words, with~\modelName's outputs, most queries can be executed almost as fast as if they are optimized with true cardinalities. 
At the 95\% quantile,~\modelName~outperforms PG, Uni-Samp, DeepDB, NeuroCard, FLAT, FactorJoin, MLP, MSCN and NNGP, in terms of P-error, by up to 26, 32, 21, 518, 1,385, 551, 1,443, 463 and 37 times, respectively. 
All these show that~\modelName~achieves large superiority over the competitors on helping PostgreSQL process queries more efficiently.

It is noteworthy that~\modelName~has clearer advantages on the STATS dataset with more complex join patterns. This reflects that our~\modelName~is able to grasp the implicit relations between complex join patterns and the underlying data significantly better than the alternatives. 
Besides,~\modelName~outperforms MLP in terms of E2E time and P-error. The main structure difference of~\modelName~and MLP is reflected on the adoption of attentions in~\modelName's modules.
This verifies the positive effects of the attention mechanisms in extracting useful information from underlying data and SQL queries.

\noindent\underline{\textbf{Estimation Accuracy.}} 
Table~\ref{tab:overall_results} also reports on the Q-error distributions of all methods.
In general,~\modelName~clearly outperforms PG on all three datasets. This verifies that the independence assumption in PG is not reasonable for some scenarios.~\modelName~still performs best among all methods in all cases. At most quantiles, \modelName results in the smallest maximum Q-error. The median Q-error of \modelName in all cases are all close to 1, the optimal value. 
At the 95\% quantile, the Q-error of~\modelName~in all cases is smaller than 10.
In contrast, none of the other methods can reach this level of performance. 
Also at the 95\% quantile, Uni-Samp, DeepDB, NeuroCard, FLAT, FactorJoin, MLP, MSCN and NNGP result in up to or more than $10^9\times$, 84$\times$, 1,643$\times$, 8,220$\times$, $10^4\times$, 3,480$\times$, $10^4\times$, 7,927$\times$, $10^4\times$ larger Q-error than that of~\modelName. 
At other quantiles, these methods are still incomparable with~\modelName.
These results show that~\modelName~is more accurate and able to discover the implicit relationships among attributes and those between queries and attributes.
Another inter-

\noindent esting thing is that smaller Q-error does not necessarily result in smaller E2E time, which is claimed in the existing work~\cite{DBLP:journals/pvldb/NegiMKMTKA21} and shown by the comparisons among NNGP, DeepDB and NeuroCard, \etc, on the Job-light dataset. An inaccurate estimate for a single sub-query tends to generate bad query plans and large execution time. It is necessary to ensure accurate estimates for all sub-queries.

\vspace{-1pt}
\noindent\underline{\textbf{Model Construction Efficiency.}}
Referring to Table~\ref{tab:overall_results}, the training cost of~\modelName~is small. It requires less than 10, 8 and 13 minutes to fine-tune its parameters for the STATS, Job-light and TPCH datasets, respectively. 
In contrast, DeepDB, NeuroCard, FLAT and MSCN consume more construction time. 
Although Uni-Samp and Factorjoin require less construction time in some cases, their E2E time, P-error and Q-error performance are much worse.
Compared to~\modelName, MLP and NNGP have simpler structures and thus they need less time to train. However, their representation abilities are not so powerful as that of~\modelName. The overwhelming advantage of~\modelName~on E2E time illustrates the necessity of a more complex structure and slightly more training time.

\vspace{-1pt}
In terms of latency, Uni-Samp achieves the worst performance. FactorJoin, MLP and MSCN require the least average time, less than 10 ms, on making an estimate. ~\modelName's latency is slightly larger but smaller than the others.
Considering the fact that executing a query on all datasets takes more than 10 seconds on average, estimation latency of less than 10 ms is not crucial in the overall picture.

\noindent\underline{\textbf{Storage Overhead.}}
The storage overhead of the query-driven methods is smaller than that of the data-driven ones in general. The query-driven methods only need to maintain a fixed number of parameters whose sizes are usually much smaller than the `data summaries' held by the data-driven methods, \emph{e.g.,} the SPN or FSPN in DeepDB and FLAT, respectively.
FactorJoin and MSCN incur the smallest memory costs.
~\modelName~consumes more memory than MLP and NNGP due to its more complex structure.
Compared to NeuroCard and FLAT,~\modelName~saves up to 75.4\% and 89.4\% memory cost on the STATS dataset, respectively. 
~\modelName~results in a little more memory cost than FLAT and DeepDB on the Job-light dataset.
Considering the fact that modern computers usually have large memories and ~\modelName~performs better in terms of E2E time, P-error and Q-error, the extra storage overhead pays off highly.

\vspace{-7pt}
\subsection{Effect of Distribution Shifting}
\label{subsec:exp_dist_shift}
\vspace{-2pt}

To investigate if~\modelName~still works well when the distribution of the underlying data greatly changes, we carry out experiments to compare the different methods on the \textsf{Dist-shift} workload. It is noteworthy that the evaluation part of the \textsf{Dist-shift} workload covers highly skewed insert statements. 
Table~\ref{tab:dist_shift_results} reports the E2E time, Q-error and P-error distribution comparisons among~\modelName, PG, NeuroCard, MSCN, NNGP and Optimal. Those competitors are chosen because they have better overall performance on the \textsf{Insert-heavy} and \textsf{Update-heavy} workloads. Due to space limit, the storage overhead, building time and latency comparisons are omitted. These results are similar to the counterparts on the other two workloads.

\begin{scriptsize}
	\begin{table}[htb]
			\vspace{-2pt}
		\centering
		\caption{Performance of methods on \textsf{Dist-shift} workloads}
		\label{tab:dist_shift_results}
		\vspace{-9pt}
		\setlength\tabcolsep{2.5pt}
		\arrayrulecolor{black}
			\begin{tabular}{   L{0.6cm} V{2}  L{0.9cm} V{2}  C{0.68cm} | C{0.6cm}  C{0.6cm}  C{0.6cm}  C{0.6cm} | C{0.45cm}  C{0.45cm}  C{0.45cm}  C{0.5cm} }
				\specialrule{1pt}{0pt}{0pt}
				\multirow{2}{*} {Data} & \multirow{2}{*}{Model} & \multirow{2}{*}{\tabincell{c}{E2E \\ Time(S)}} & \multicolumn{4}{c |}{Q-error} & \multicolumn{4}{c}{P-error}  \\ \hhline{*{3}{|~}*{4}{|-}|} \hhline{*{7}{|~}*{4}{|-}|} 
				& & & 50\% & 90\% & 95\% & 99\% & 50\% & 90\% & 95\% & 99\%  \\ \specialrule{1pt}{0pt}{0pt}
				\multirow{7}{*} {STATS} 
				& PG & \cellcolor{grey}$8,432$ & $189$ & $1.4$$\cdot$$10^5$ & $1.1$$\cdot$$10^6$ & $1.9$$\cdot$$10^7$ & $2.60$ & $25.50$ & $42.65$ & $300$   \\  
				& Uni-Samp & \cellcolor{green2}$7,524$ & $\boldsymbol{1.32}$ & >$10^{10}$ & >$10^{10}$ & >$10^{10}$ & $1.19$ & $12.62$ & $38.63$ & $87.59$ \\
				& NeuroCard & $27,252$ & $14.30$ & $996$ & $5,051$ & $3.9$$\cdot$$10^5$ & $2.08$ & $25.69$ & $108$ & $3,318$   \\ 
				& MSCN &  $26,697$ & $20.08$ & $2,802$ & $15,576$ & $4.9$$\cdot$$10^5$ & $2.75$ & $46.51$ & $465$ & $83,452$ \\
				& NNGP & $10,537$ & $9.47$ & $785$ & $4,370$ & $2.4$$\cdot$$10^5$ & $1.45$ & $7.19$ & $25.19$ & $3,318$ \\
				& \modelName & \cellcolor{green4}$\boldsymbol{6,876}$ & $1.40$ & $\boldsymbol{5.46}$ & $\boldsymbol{11.75}$ & $\boldsymbol{118.36}$ & $\boldsymbol{1.07}$ & $\boldsymbol{1.38}$ & $\boldsymbol{1.71}$ & $\boldsymbol{11.18}$ \\ 
				& Optimal& \cellcolor{aqua}$6,770$ & $1$ & $1$ & $1$ & $1$ & $1$ & $1$ & $1$ & $1$\\ \specialrule{1pt}{0pt}{0pt}
				\multirow{7}{*} {\tabincell{l}{Job-\\light}} 
				& PG & \cellcolor{grey}$5,608$ & $1.58$ & $6.12$ & $14.22$ & $76.30$ & $1.10$ & $1.32$ & $1.48$ & $2.89$  \\  
				& Uni-Samp & $5,906$ & $1.44$ & $3.46$ & $10.78$ & >$10^{10}$ & $1.16$ & $1.68$ & $1.90$ & $3.28$\\
				& NeuroCard & $17,720$ & $14.15$ & $822$ & $4,799$ & $3.5$$\cdot$$10^5$ & $1.81$ & $12.40$ & $37.25$ & $144$ \\ 
				& MSCN &  $25,157$ & $28.54$ & $1,592$ & $5,261$ & $99,178$ & $2.06$ & $15.01$ & $27.77$ & $84.35$ \\
				& NNGP & $11,698$ & $4.6$$\cdot$$10^8$ & >$10^{10}$ & >$10^{10}$ & >$10^{10}$ & $3.50$ & $7.90$ & $11.24$ & $23.38$\\
				& \modelName & \cellcolor{green4}$\boldsymbol{4,763}$ & $\boldsymbol{1.16}$ & $\boldsymbol{2.23}$ & $\boldsymbol{4.34}$ & $\boldsymbol{16.22}$ & $\boldsymbol{1.07}$ & $\boldsymbol{1.24}$ & $\boldsymbol{1.32}$ & $\boldsymbol{1.77}$\\ 
				& Optimal& \cellcolor{aqua}$4,708$ & $1$ & $1$ & $1$ & $1$ & $1$ & $1$ & $1$ & $1$\\ \specialrule{1pt}{0pt}{0pt}
				\multirow{7}{*} {TPCH} 
				& PG & \cellcolor{grey}$3,377$ & $1.23$ & $3.94$ & $5.85$ & $10.28$ & $1.10$ & $1.39$ & $1.64$ & $3.06$  \\  
				& Uni-Samp & >$30,000$ & $1.29$ & $3.24$ & >$10^{10}$ & >$10^{10}$ & $1.79$ & $2.28$ & $62.54$ & $301$ \\
				& NeuroCard  & $11,362$ & $44.65$ & $9,151$ & $34,541$ & $2.8$$\cdot$$10^5$ &  $1.60$ & $15.10$ & $25.94$ & $50.68$  \\ 
				& MSCN & $7,062$ & $41.62$ & $6,571$ & $28,837$ & $2.4$$\cdot$$10^5$ & $3.64$ & $9.17$ & $22.00$ & $202$ \\
				& NNGP & $18,244$ & $1.4$$\cdot$$10^9$ & >$10^{10}$ & >$10^{10}$ & >$10^{10}$ & $5.54$ & $46.30$ & $59.12$ & $99.12$ \\
				& \modelName & \cellcolor{green4}$\boldsymbol{3,230}$ & $\boldsymbol{1.18}$ & $\boldsymbol{2.66}$ & $\boldsymbol{4.46}$ & $\boldsymbol{11.23}$ & $\boldsymbol{1.07}$ & $\boldsymbol{1.23}$ & ${1.29}$ & $\boldsymbol{1.57}$\\ 
				& Optimal& \cellcolor{aqua}$3,226$ & $1$ & $1$ & $1$ & $1$ & $1$ & $1$ & $1$ & $1$ \\ \specialrule{1pt}{0pt}{0pt}
			\end{tabular}
		\end{table} 
\end{scriptsize}
 \vspace{-1pt}

Referring to Table~\ref{tab:dist_shift_results}, the overall Q-error and P-error performance of all methods on the \textsf{Dist-shift} workloads are worse than the counterparts on the \textsf{Insert-heavy} workloads. 
This is reasonable because compared to the data upon which these methods are built, the distribution of the underlying data is greatly shifted when testing queries are executed.
Nevertheless,~\modelName~still achieves the best E2E time on all datasets. It needs up to 18.5\% less time than PG to execute all testing queries. Its E2E time on all three workloads is close to the optimal results and much smaller than that of other competitors.
In terms of Q-error and P-error,~\modelName~also performs best. At the 95\% quantile, the largest Q-error and P-error of~\modelName~are 11.75 and 1.71, respectively. In contrast, the other competitors' Q-error and P-error are at least $2.48$ and $1.44$ times larger, respectively.
All these demonstrate that~\modelName~is less sensitive to the distribution shifting and able to make accurate estimates even when the distributions of the underlying data changes significantly.

\vspace{-5pt}
\section{Related Work}
\label{sec:related_work}

\noindent\textbf{Data-driven Cardinality Estimators.}
%
Data-driven methods aim to describe the underlying data with statistical or machine learning models.
The simple yet efficient 1-D Histogram~\cite{DBLP:conf/sigmod/SelingerACLP79} is used in many well-known DBMS like PostgreSQL. It assumes all attributes are mutually independent and maintains a 1-D (cumulative) histogram for each attribute. 
To address the problem of unreasonable independence assumption, M-D Histogram based methods~\cite{DBLP:conf/sigmod/DeshpandeGR01,DBLP:conf/sigmod/GunopulosKTD00,DBLP:conf/vldb/PoosalaI97,DBLP:conf/cascon/WangS03} build multi-dimensional histograms to model attribute dependency. 
Although such methods improve the accuracy, the decomposition of the joint attributes is still lossy. Also, they hardly work for queries with complex joins.
Sampling-based methods~\cite{DBLP:conf/sigmod/ChaudhuriMN99,DBLP:conf/sigmod/ChenY17,DBLP:conf/cidr/KipfKRLBK19,DBLP:conf/sigmod/QiuWYLWZ21,DBLP:conf/sigmod/0001WYZ16} address join queries but they risk high variance and sampling failure when the data distribution or query is complex.
Bayesian network (BN) based methods~\cite{DBLP:journals/tit/ChowL68,DBLP:conf/sigmod/GetoorTK01,DBLP:journals/pvldb/TzoumasDJ11} use a directed acyclic graph to model the dependence among attributes, assuming that each attribute is conditionally independent of the remaining attributes given its parents' distributions.
BayesCard~\cite{DBLP:journals/corr/abs-2012-14743} revitalizes BN using probabilistic programming to improve its inference and model construction speed.
Recently, machine learning techniques are adopted in data-driven methods.
Deep autoregressive models are adopted in Naru~\cite{DBLP:journals/pvldb/YangLKWDCAHKS19} and NeuroCard~\cite{DBLP:journals/pvldb/YangKLLDCS20} to decompose the joint distribution of attributes to a product of conditional distributions.
DeepDB~\cite{DBLP:journals/pvldb/HilprechtSKMKB20} is built upon Sum-Product Network (SPN)~\cite{DBLP:conf/uai/PoonD11} which approximates the joint distribution using several local and simple PDFs.
FLAT~\cite{DBLP:journals/pvldb/ZhuWHZPQZC21} improves SPN by adopting a factorize-split-sum-product network (FSPN)~\cite{DBLP:journals/corr/abs-2011-09020} to adaptively decompose the joint distribution according to the attribute dependence level.

\noindent\textbf{Query-driven Cardinality Estimators.} Such estimators focus on modeling the relationships between queries and their true cardinalities.
The feedbacks of past queries are utilized to correct and self-tune histograms~\cite{DBLP:conf/sigmod/BrunoCG01,DBLP:journals/isci/FuchsHL07,DBLP:journals/tkde/KhachatryanMSB15,DBLP:conf/icde/SrivastavaHMKT06} and update statistical summaries~\cite{DBLP:conf/vldb/StillgerLMK01,DBLP:journals/pvldb/WuJAPLQR18}. 
LW-XGB and LW-NN~\cite{DBLP:journals/pvldb/DuttWNKNC19} formulate the cardinality estimation as a regression problem and apply gradient boosted trees and neural networks for regression, respectively.
UAE-Q~\cite{DBLP:conf/cidr/WuYYZHLLZZ22} applies the deep auto-regression models and differentiable progressive sampling via the Gumbel-Softmax trick to learn hidden information from queries.
The KDE-based join estimators~\cite{DBLP:journals/pvldb/KieferHBM17} combine kernel density estimation (KDE) with a query-driven tuning mechanism to estimate multivariate probability distributions of a relation and cardinalities of joins.
Fauce~\cite{DBLP:journals/pvldb/LiuD0Z21} and NNGP~\cite{DBLP:conf/sigmod/ZhaoYHLZ22} assume a query's cardinality follows a Gaussian distribution and adopt Deep Ensemble~\cite{DBLP:conf/nips/Lakshminarayanan17} and neural network Gaussian process~\cite{DBLP:conf/iclr/LeeBNSPS18} to predict the distribution's mean and variance.
%


A few works also consider both data and SQL queries.
Wu~\emph{et. al.}~\cite{DBLP:conf/sigmod/WuC21} propose a unified deep autoregressive model utilizing both data as unsupervised information and query workload as supervised information.
Kipf~\emph{et. al.}~\cite{DBLP:conf/cidr/KipfKRLBK19} concatenate basic relation information and query features together and use a multi-set convolutional neural network to process them and make estimates. 
Negi~\emph{et. al.}~\cite{DBLP:journals/pvldb/NegiWKTMMKA23} propose techniques to build sample tables on the join keys and use neural networks to extract information from queries. 
However, these methods require either samples over joined attributes which results in high sampling overhead, or unrenewable featurization/samples over static data which are not applicable for dynamic database.
Also, compared to~\modelName, these methods do not explain the links among data, queries and true cardinalities.
Besides, Negi's work~\cite{DBLP:journals/pvldb/NegiWKTMMKA23} supports PK-FK joins only.
Dutt \emph{et. al.}~\cite{DBLP:journals/pvldb/DuttWNKNC19} use neural networks and tree-based ensembles to extract information from data and queries to solve the problem of selectivity estimation on a single relation. Thus, their approach does not support joins.
In addition, Han~\emph{et. al.}~\cite{DBLP:journals/pvldb/HanWWZYTZCQPQZL21} propose an end-to-end evaluation benchmark for cardinality estimators. Our used benchmark is an improved version of it. Sun~\emph{et. al.}~\cite{DBLP:journals/pvldb/SunZSLT21} make a comprehensive comparison of the existing cardinality estimators.


\vspace{-5pt}
\section{Conclusion and Future Work}
\label{sec:conclusion}

In this work, we design~\modelName, a versatile learned cardinality estimation model, that makes accurate and high-quality estimates for SQL queries. Based on two delicate methods to featurize the underlying database data and the SQL queries, respectively,~\modelName~adopts the attention mechanisms in its two modules to understand the
implicit relations between data and queries.
The self-attention layer in the data-encoder module figures out the links among all database attributes.
The query-analyzer takes the input of the query featurization and the output of the data-encoder, and puts attention on the more important parts of the data. Extensive experimental results show that~\modelName~clearly outperforms the state-of-the-art alternatives in terms of multiple evaluation metrics. 

For future work, it is interesting to  extend~\modelName~to more general aggregate analytic queries by replacing $\mathtt{COUNT}(*)$ with other aggregate functions.
Also, it is relevant to explore if better data and query featurization methods exist.
Moreover, it makes sense to use other types of attention functions in~\modelName.

\balance

\if\releaseversion0
\bibliographystyle{ACM-Reference-Format}
\bibliography{ref}
\fi

\if\releaseversion1
\if\includeAppendix1

\fi

\if\includeAppendix1
\begin{figure*}
	\begin{center}
		\Huge\sffamily\bfseries Appendix~\label{Appendix}
	\end{center}
\end{figure*}

\appendix
\newpage

\noindent
Due to space limit, we present some minute details and relatively less important experimental results and analyses in this appendix.

\section{Multi-head Attention}
\label{sec:multihead_attn}
To enhance the representation ability of our~\modelName, we adopt a multi-head attention mechanism~\cite{DBLP:conf/nips/VaswaniSPUJGKP17} in both modules instead of performing a single function. Fig.~\ref{fig:multi_head_attn} depicts its details.
\setlength{\textfloatsep}{1pt}
\setlength{\columnsep}{1pt}
\setlength{\intextsep}{1pt}
\begin{figure}[htb]
	\centering
	\vspace{-2pt}
	\if\includeAppendix1
	\includegraphics[width=0.28\textwidth, trim={5mm 5mm 5mm 5mm},clip]{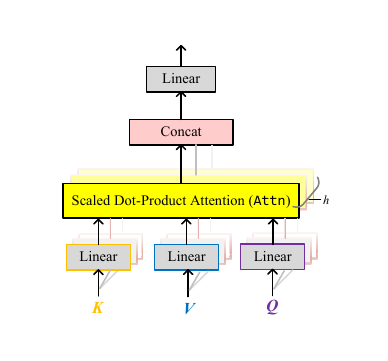}
	\else
	\includegraphics[width=0.28\textwidth, trim={5mm 5mm 5mm 5mm},clip]{multi_head_attn.eps}
	\fi
	\vspace{-12pt}
	\caption{Multi-head Attention, adapted from~\cite{DBLP:conf/nips/VaswaniSPUJGKP17}.}
	\label{fig:multi_head_attn}
\end{figure}

A multi-head attention projects the queries, keys and values multiple times with $h$ different linear projections, and then executes the attention function $\mathsf{Attn}$ with the input of the projected queries, keys and values in parallel. Compared to a single attention head, it could jointly attend to information from different projected spaces and thus is more powerful. Below is its analytical expressions:
\begin{flalign}\nonumber
	\text{MutliHead}(\boldsymbol{Q, K, V}) &= \text{concat}(\text{head}_1, \cdots, \text{head}_h) \boldsymbol{W}^{M}, \\ \nonumber
	\text{where~head}_i &= \mathsf{Attn}(\boldsymbol{Q}\boldsymbol{W}_{i}^{\boldsymbol{Q}}, \boldsymbol{K}\boldsymbol{W}_{i}^{K}, \boldsymbol{V}\boldsymbol{W}_{i}^{V})
\end{flalign}
Above, $\boldsymbol{W}_{i}^{Q} \in \mathbb{R}^{d_k \times d_m}$, $\boldsymbol{W}_{i}^{Q} \in \mathbb{R}^{d_k \times d_m}$ and $\boldsymbol{W}_{i}^{Q} \in \mathbb{R}^{d_v \times d_m}$ project each query, key and value vector into $\mathbb{R}^{d_m}$ space respectively; $\boldsymbol{W}^{M} \in \mathbb{R}^{d_m \times d_v}$ projects the output of weighted values back to $\mathbb{R}^{d_v}$. 
In our experiments, the value of $h$ is set to 8, following the settings in~\cite{DBLP:conf/nips/VaswaniSPUJGKP17}.

It is noteworthy that the choice of the attention function $\mathsf{Attn}$ is not unique and there may be other type of $\mathsf{Attn}$-integrated attention mechanism besides the multi-head attention. The attention layers can be flexibly designed depending on situations.

\begin{figure*}[htb]
	\centering
	\if\includeAppendix1
	\includegraphics[width=1\textwidth, trim={5mm 6mm 1mm 3mm},clip]{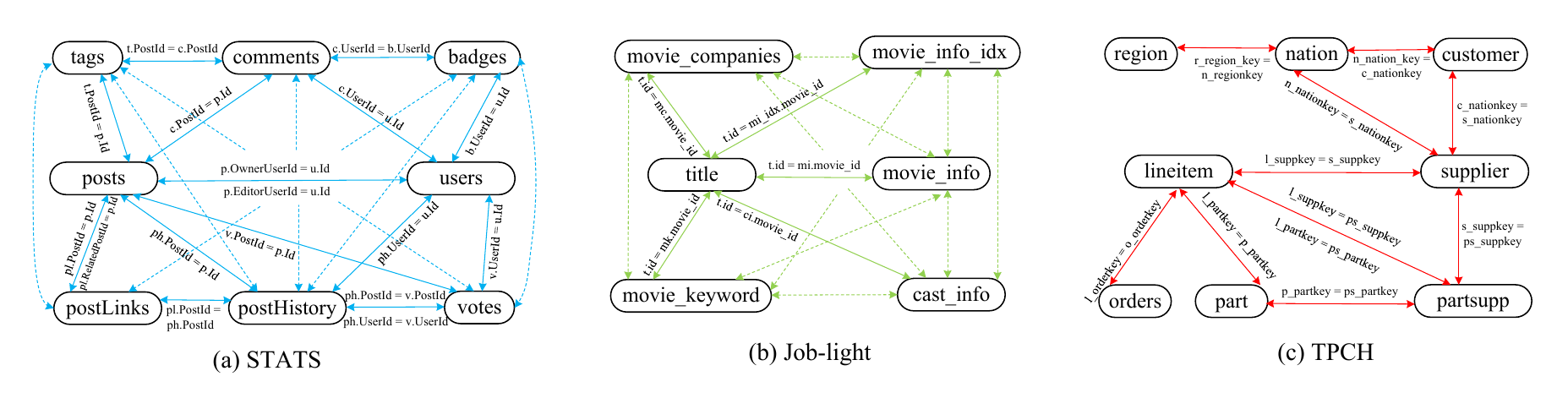}
	\else
	\includegraphics[width=1\textwidth, trim={5mm 6mm 1mm 3mm},clip]{joinRelations.eps}
	\fi
	\vspace{-22pt}
	\caption{Joins among relations in three datasets.}
	\label{fig:joins}
	\vspace{-5pt}
\end{figure*}

\section{An Improved Benchmark}
\label{sec:benchmark_usage}

\noindent The \emph{CardEst} benchmark~\cite{DBLP:journals/pvldb/HanWWZYTZCQPQZL21} provides a way to integrate external cardinality estimators into the built-in query optimizer of PostgreSQL. 
In particular, given a SQL query to be executed, PostgreSQL's query optimizer needs to get the estimated cardinalities of a series of sub-queries in a fixed order. 
By enabling a knob, the benchmark provides the function of reading the cardinality estimates from a given file instead of using the estimates by PostgreSQL's built-in estimator.
Thus, if we know what the sub-queries are and feed the corresponding cardinality estimates accessed through external estimators, we can directly compare the quality of their generated query plans by watching the end-to-end query execution time.

The idea above is simple and appealing. However, the original benchmark~\cite{DBLP:journals/pvldb/HanWWZYTZCQPQZL21} cannot produce correct sub-queries for queries on the dynamic workloads and with complex join predicates. In other words, it works on very limited cases only. 
These drawbacks motivate us to improve the benchmark to support dynamic workloads and more general join schema. 
Our improved benchmark~\cite{codes_url} implements correct sub-queries generations function and inherits the cardinality estimates reading function. 
We have validated it with numerous queries on the STATS, Job-light and TPCH datasets.

\noindent\textbf{\underline{Usage guidance.}} Our benchmark mainly provides two functions: sub-queries generation and cardinality estimates replacement. We leave multiple knobs to let users decide when to generate sub-queries and which method's estimation results used to generate the query plans for the queries on a workload $\boldsymbol{W}$.

\emph{1) Sub-queries generation.} This function is related with only one knob: \emph{print\_sub\_queries}. Given a query, the optimizer needs the cardinality estimations of the sub-queries of two types: the \emph{single} sub-queries only involve single tables whereas the \emph{join} sub-queries cover join conditions.
By setting the value of the knob to \textsf{True} and executing the \texttt{Explain} statement for each query on $\boldsymbol{W}$, two files will be generated in the data directory of PostgreSQL: `\texttt{single\_sub\_queries.txt}' and `\texttt{join\_sub\_queries.txt}'. They record the above two types of sub-queries, respectively. Each line of either file is a sub-query of a query $\sqlq$ on $\boldsymbol{W}$. Also, the line will include the appearance ranking of $\sqlq$ among all queries. This information help users know the ties between queries and sub-queries.

\emph{2) Cardinality estimates replacement.}
After estimating the cardinalities of the sub-queries in both files using an external method, \emph{e.g.}~\modelName, we can `inject' them into PostgreSQL to replace the built-in results. 
First, we need to save the estimation results into two files with each file corresponds to one type of sub-queries, and copy them into the data directory of PostgreSQL.
Then, two knobs, namely \emph{read\_single\_cards} and \emph{read\_join\_cards}, are supposed to be turned on. Meanwhile, we should set the configuration parameters \emph{single\_cards\_fname} an \emph{join\_cards\_fname} to be the names of the above two files, respectively.
After these settings, we can run the workload in the usual way. It is noteworthy that the query optimizer will use the cardinality estimates by the external method to generate the query plans.
Suppose the files for the single and join sub-queries are named `\texttt{single\_cards.txt}' and `\texttt{join\_cards.txt}', respectively. The setting statements for the knobs and configuration parameters are shown as follows.

\begin{lstlisting}
	SET read_single_cards=true;
	SET read_join_cards=true;
	SET single_cards_fname='single_cards.txt';
	SET join_cards_fname='join_cards.txt';
\end{lstlisting}

Usually, PostgreSQL's built-in estimator is able to give sufficiently accurate cardinality estimates for the single sub-queries. Thus, we only need to estimate the cardinalities of the join sub-queries, turn on the knob \emph{read\_join\_cards} and set the parameter \emph{join\_cards\_fname} in most cases.

Please also refer to our github repository~\cite{codes_url} for more details.

\section{Additional Experimental Results}
\label{sec:additional_exp}

This section presents the additional experimental results that cannot be put in the main body of the paper due to the space limit.

\subsection{Join Information among Relations}
Fig.~\ref{fig:joins} shows the join information among relations in the STATS, Job-light and TPCH datasets. For clearity, only part of the joins are exhibited.

\subsection{Performance on Static Workloads}
\label{subsec:exp_static_workload}

We also conduct experiments to compare all methods on the static workloads that consist of query statements only. In particular, we use the whole STATS, Job-light and TPCH datasets to build the data-driven methods. On top of the whole dataset, the training queries and sub-queries as well as their true cardinalities are used as the training and validation data for the query-driven methods. Accordingly, all data-driven and query-driven methods estimate the cardinalities for the testing sub-queries. These estimates are in turn used to generate the E2E evaluation results, Q-error, \etc. Table~\ref{tab:static_workload_results} reports the relevant comparative results.

\setlength{\textfloatsep}{\oldtextfloatsep}
\setlength{\columnsep}{\oldcolumnsep}
\setlength{\intextsep}{\oldintextsep}
\begin{scriptsize}
\begin{table}[htb]
		\vspace{-3pt}
	\centering
	\caption{Performance of methods on \textsf{static} workloads}
	\label{tab:static_workload_results}
	\vspace{-5pt}
	\setlength\tabcolsep{2.5pt}
	\arrayrulecolor{black}
		\begin{tabular}{  L{0.6cm} V{2} l V{2}  C{0.8cm} | c  c  c  c | c |  C{0.9cm} | C{0.6cm}  }
			\specialrule{1pt}{0pt}{0pt}
			\multirow{2}{*} {Data} & \multirow{2}{*}{Model} & \multirow{2}{*}{\tabincell{c}{E2E \\ Time(S)}} & \multicolumn{4}{c |}{Q-error} & \multirow{2}{*}{\tabincell{c}{Size \\ (MB)}} & \multirow{2}{*}{\tabincell{c}{Building \\ Time(Min)} } &  \multirow{2}{*}{\tabincell{c}{Latency \\ (ms)}} \\ \hhline{*{3}{|~}*{4}{|-}|} 
			& & & 50\% & 90\% & 95\% & 99\% &  &  &   \\ \specialrule{1pt}{0pt}{0pt}
			\multirow{8}{*} {STATS} 
			& PG & \cellcolor{grey}$12,777$ & $1.80$ & $21.84$ & $106$ & $7,950$ & - & - & -  \\  
			& Uni-Samp & $15,397$ & $\boldsymbol{1.33}$ & $6.64$ & >$10^{10}$ & >$10^{10}$ & $4.44$ & $\boldsymbol{0.02}$ & $413$ \\
			& NeuroCard & $19,847$ & $2.91$ & $192$ & $1,511$ & $1.5\cdot10^5$ & $121.88$ & $27.77$ & $40.16$  \\ 
			& MLP & \cellcolor{green4}$7,823$ & $1.58$ & $4.62$ & $9.22$ & $76.97$ & $8.52$ & $3.10$ & $3.10$ \\
			& MSCN & $15,162$ & $3.85$ & $39.56$ & $99.81$ & $1,273$ & $\boldsymbol{1.61}$ & $12.41$  & $\boldsymbol{0.79}$ \\
			& NNGP & $20,181$ & $8.10$ & $694$ & $3,294$ & $2.3$$\cdot$$10^5$ & $9.56$ & $0.68$ & $21.28$  \\
			& \modelName & \cellcolor{green4}$\boldsymbol{7,704}$ & $1.47$ & $\boldsymbol{4.86}$ & $\boldsymbol{9.11}$ & $\boldsymbol{56.98}$ & $22.31$ & $6.44$ & $8.64$ \\ 
			& Optimal& \cellcolor{aqua}$7,622$ & $1$ & $1$ & $1$ & $1$ & - & - & - \\ \specialrule{1pt}{0pt}{0pt}
			\multirow{8}{*} {\tabincell{l}{Job-\\light}} 
			& PG & \cellcolor{grey}$19,820$ & $1.70$ & $9.41$ & $16.25$ & $50.19$& - & - & -  \\  
			& Uni-Samp & $19,146$ & $\boldsymbol{1.32}$ & $3.31$ & $6.00$ & >$10^{10}$ & $31.83$ & $\boldsymbol{0.48}$ &  $187$ \\
			& NeuroCard & \cellcolor{green4}$18,153$ & $1.37$ & $4.16$ & $6.35$ & $15.22$ & $49.96$ & $9.83$ & $16.76$  \\ 
			& MLP & \cellcolor{green4}$18,413$ & $1.40$ & $3.78$ & $7.00$ & $76.30$ & $6.19$ & $1.83$ & $2.25$ \\
			& MSCN & $24,829$ & $10.94$ & $200$ & $396$ & $2,741$ & $\boldsymbol{1.56}$ & $31.68$ & $\boldsymbol{1.04}$ \\
			& NNGP & >$30,000$ & $6.66$ & $121$ & $414$ & $1.5$$\cdot$$10^5$ & $32.44$& $\boldsymbol{0.88}$ & $30.35$  \\
			& \modelName & \cellcolor{green4}$\boldsymbol{18,015}$ & $1.44$ & $\boldsymbol{3.23}$ & $\boldsymbol{5.75}$ & $\boldsymbol{47.28}$ & $22.25$ & $5.88$ & $7.32$ \\ 
			& Optimal& \cellcolor{aqua}$17,939$ & $1$ & $1$ & $1$ & $1$ & - & - & -\\ \specialrule{1pt}{0pt}{0pt}
			\multirow{8}{*} {TPCH} 
			& PG & \cellcolor{grey}$12,217$ & $1.23$ & $3.95$ & $6.22$ & $11.33$ & - & - & -  \\  
			& Uni-Samp & $19,319$ & $1.16$ & $3.09$ & >$10^{10}$ & >$10^{10}$  & $30.95$ & $\boldsymbol{0.14}$ & $26.24$  \\
			& NeuroCard & \cellcolor{green2}$12,020$ & $\boldsymbol{1.09}$ & $2.97$ & $395$ & $450$& $850.53$ & $44.39$ & $35.98$  \\ 
			& MLP & \cellcolor{green4}$8,957$ & $1.44$ & $2.50$ & $3.79$ & $10.91$ & $8.63$ & $2.68$ & $3.67$ \\
			& MSCN & \cellcolor{green2}$11,893$ & $4.29$ & $36.76$ & $83.76$ & $321$ & $\boldsymbol{1.58}$ & $24.17$ & $\boldsymbol{0.80}$ \\
			& NNGP & $13,183$ &$31.89$ & $1,052$ & $2,104$ & $11,811$ & $32.44$ & $0.91$ & $39.66$ \\
			& \modelName & \cellcolor{green4}$\boldsymbol{8,717}$ & $1.24$ & $\boldsymbol{2.36}$ & $\boldsymbol{4.26}$ & $\boldsymbol{10.73}$ & $22.34$ & $7.15$ & $10.31$ \\ 
			& Optimal& \cellcolor{aqua}$8,706$ & $1$ & $1$ & $1$ & $1$ & - & - & -  \\ \specialrule{1pt}{0pt}{0pt}
		\end{tabular}
\end{table} 
\end{scriptsize}
	
As shown in Table~\ref{tab:static_workload_results}, the end-to-end performance of cardinality estimators on the static workloads is different from the counterparts on the dynamic workloads.
Overall, the performance of the methods on static workloads is better than that on the dynamic workloads. Nevertheless,~\modelName~still performs best among all methods.
~\modelName~results in up to 1.66, 2.00, 2.58, 1.97 and at least 3 times faster query execution than PG, Uni-Samp, NeuroCard, MSCN and NNGP, respectively. The E2E time of~\modelName~is only at most 1.1\% larger than that of Optimal. 
Next, at most quantiles, ~\modelName's Q-error is smaller than that of the others on all datasets. 
At the 95\% quantile,~\modelName~achieves up to or more than $11.64\times$, $10^9\times$, $166\times$, $77\times$ and $494\times$ smaller Q-error compared to PG, Uni-Samp, NeuroCard, MSCN and NNGP, respectively. 
~\modelName~is not the best in terms of training time, latency and memory cost. However,~\modelName~incurs comparable results with the competitors. 
Considering the much better E2E time and estimation accuracy that~\modelName~achieves, the slightly extra overhead is acceptable.

\noindent\textbf{\underline{\modelName~vs. MLP on static workloads.}}
Referring to Table~\ref{tab:static_workload_results}, MLP achieves  E2E time and Q-error comparable with the counterparts of~\modelName, and even less storage overhead, latency and training time.
The reason behind is that~\modelName~is almost equivalent to MLP when processing queries on static workloads, where the DB states for the queries are constant. 
In this case, only one element exists in the input sets of keys, values and queries of the self-attention layers in the data-encoder module. This is the same to the three sets of the attention layers in the query-analyzer module. 
Consequently, both attention layers make almost no effects, and the whole~\modelName~network degenerates to an MLP.
Therefore, if there is no underlying data change, we can simply train an MLP to estimate cardinalities.

\subsection{Effect of $\dx$}
\label{subsec:dx_effect}

Parameter $\dx$ is used to control the amount of distribution information covered by data featurizations. A larger $\dx$ usually implies more informative featurizations but results in more storage and latency overhead. To investigate the effect of $\dx$, we build four~\modelName versions with different $\dx$ values and observe their performance on the \textsf{Insert-heavy} workload of the STATS dataset. The results on the other datasets and other workloads are similar and thus omitted.
Table~\ref{tab:dx_effects} shows the results of~\modelName~with different values of $\dx$. 
As $\dx$'s value increases,~\modelName~is able to make more accurate estimates. Meanwhile, the building time does not changes much. The storage overhead and latency increase slightly. 
In particular, when the value of $\dx$ increases from $10$ to $40$, the memory cost of \modelName increases by only $0.51$ MB.
Also, the value of $\dx$ does not influence the cost of updating DB states when an data manipulation statement is executed. Thus, we suggest to use a larger $\dx$ to build the DB states within affordable memory and latency costs.
\begin{scriptsize}
	\begin{table}[htb]
		\centering
		\caption{The effects of the hyperparameter $\dx$}
		\label{tab:dx_effects}
		\vspace{-10pt}
		\setlength\tabcolsep{2.5pt}
		\arrayrulecolor{black}
			\begin{tabular}{  l | c  c  c  c | c | c | c  }
				\specialrule{1pt}{0pt}{0pt}
				\multirow{2}{*}{$\dx$} & \multicolumn{4}{c |}{Q-error} & \multirow{2}{*}{\tabincell{c}{Size \\ (MB)}} & \multirow{2}{*}{\tabincell{c}{Building \\ Time(Min)} } &  \multirow{2}{*}{\tabincell{c}{Latency \\ (ms)}} \\ \hhline{*{1}{|~}*{4}{|-}|} 
				& 50\% & 90\% & 95\% & 99\% &  &  &   \\ \specialrule{1pt}{0pt}{0pt}
				10 & $1.26$ & $5.52$ & $13.57$ & $78.16$& $\boldsymbol{22.15}$ &$9.47$ & $\boldsymbol{8.08}$ \\  
				20 & $1.33$ & $5.49$ & $12.65$ & $65.52$ & $22.21$ & $\boldsymbol{8.83}$ & $8.11$ \\
				40 & $1.29$ & $5.07$ & $11.54$ & $\boldsymbol{62.52}$ & $22.31$ & $9.25$ & $8.25$ \\
				80 & $\boldsymbol{1.24}$ & $\boldsymbol{5.00}$ & $\boldsymbol{10.76}$ & $64.25$ & $22.56$ & $9.66$ & $8.78$ \\\specialrule{1pt}{0pt}{0pt}
			\end{tabular}
		\end{table} 
	\end{scriptsize}

\subsection{Hyperparameter Studies}
\label{subsec:exp_param_studies}
To study the effects of more hyperparameters, we build different~\modelName~versions and observe their performance. Similarly, we only show the comparison results on the \textsf{Insert-heavy} workload of the STATS dataset. The results on the other datasets and other workloads are similar and thus omitted.

\noindent\underline{\textbf{Effects of $\nenc$ and $\nana$.}} 
$\nenc$ and $\nana$ are the numbers of stacked multi-head attention layers in~\modelName's data-encoder and query-analyzer modules, respectively. To investigate if they affect~\modelName's performance, we train a series of~\modelName~with different $\nenc$ and $\nana$ values. Table~\ref{tab:nenc_nana_effects} reports the comparison results.
\begin{scriptsize}
\begin{table}[htb]
	\centering
	\caption{The effects of the hyperparameters $\nenc$ and $\nana$}
	\label{tab:nenc_nana_effects}
	\vspace{-6pt}
	\setlength\tabcolsep{2.5pt}
	\arrayrulecolor{black}
		\begin{tabular}{  c  c | c  c  c  c | c | c | c  }
			\specialrule{1pt}{0pt}{0pt}
			\multirow{2}{*}{$\nenc$} & \multirow{2}{*}{$\nana$} & \multicolumn{4}{c |}{Q-error} & \multirow{2}{*}{\tabincell{c}{Size \\ (MB)}} & \multirow{2}{*}{\tabincell{c}{Building \\ Time(Min)} } &  \multirow{2}{*}{\tabincell{c}{Latency \\ (ms)}} \\ \hhline{*{2}{|~}*{4}{|-}|} 
			& & 50\% & 90\% & 95\% & 99\% &  &  &   \\ \specialrule{1pt}{0pt}{0pt}
			2 & 2 & $1.37$ & $6.33$ & $13.69$ & $90.04$ & $\boldsymbol{10.36}$ & $\boldsymbol{7.12}$ & $\boldsymbol{6.13}$ \\
			2 & 4 & $1.32$ & $5.95$ & $11.77$ & $67.02$ & $16.33$ & $8.43$ & $7.47$  \\  
			4 & 2 & $1.35$ & $5.88$ & $11.27$ & $78.26$ & $16.33$ & $7.69$ & $7.62$  \\
			4 & 4 & $\boldsymbol{1.29}$ & $5.07$ & $11.54$ & $\boldsymbol{62.52}$ & $22.31$ &$ 9.25$ & $8.25$ \\
			4 & 6 & $1.32$ & $5.57$ & $11.01$ & $73.36$ & $28.29$ & $10.17$ & $8.80$ \\
			6 & 4 & $1.45$ & $5.59$ & $11.60$ & $69.10$ & $28.29$ & $9.05$ & $8.92$ \\
			6 & 6 & $\boldsymbol{1.29}$ & $\boldsymbol{4.74}$ & $\boldsymbol{10.45}$ & $75.22$ & $34.26$ & $9.92$ & $9.19$ \\\specialrule{1pt}{0pt}{0pt}
		\end{tabular}
	\vspace{3pt}
\end{table} 
\end{scriptsize}

As shown in Table~\ref{tab:nenc_nana_effects}, the storage overhead of \modelName is clearly influenced by $\nenc$ and $\nana$. When the values of $\nenc$ and $\nana$ get larger, \modelName will also result in larger memory costs.
However, the Q-error performance of \modelName is not necessarily positively correlated to the numbers of attention layers in its two modules.
When the values of $\nenc$ and $\nana$ are set to 2, a smaller number,~\modelName~achieves the worst estimation accuracy. When both $\nenc$ and $\nana$ equal 4 or 6,~\modelName~has the overall best Q-error performance. However, larger $\nenc$ or $\nana$ will result in larger storage and latency overhead. Thus, we set the values of both parameters to 4.
	
\noindent\underline{\textbf{Effects of join condition featurization ways.}}
In Section~\labelQueryFeat, we mention that compared to simply featurize whether each join predicate appears in the SQL query, our way of additionally featurizing the relations involved in joins are more compact, informative and helpful to the estimations. To verify this claim, we carry out an ablation study by building two~\modelName versions with different join predicate featurization ways and observing their performance. The comparison results are reported in Table~\ref{tab:join_cond_featurization_way_effects}.
	
\begin{scriptsize}
\begin{table}[htb]
	\centering
	\caption{The effects of  join predicate featurization ways}
	\label{tab:join_cond_featurization_way_effects}
	\vspace{-8pt}
	\setlength\tabcolsep{2.5pt}
	\arrayrulecolor{black}
		\begin{tabular}{  l  | c  c  c  c | c | c | c  }
			\specialrule{1pt}{0pt}{0pt}
			\multirow{2}{*}{\tabincell{l}{Featurization\\way}} & \multicolumn{4}{c |}{Q-error} & \multirow{2}{*}{\tabincell{c}{Size \\ (MB)}} & \multirow{2}{*}{\tabincell{c}{Building \\ Time(Min)} } &  \multirow{2}{*}{\tabincell{c}{Latency \\ (ms)}} \\ \hhline{*{1}{|~}*{4}{|-}|} 
			& 50\% & 90\% & 95\% & 99\% &  &  &   \\ \specialrule{1pt}{0pt}{0pt}
			Simple~\cite{DBLP:conf/sigmod/ZhaoYHLZ22} & $1.36$ & $5.78$ & $\boldsymbol{11.48}$ & $73.85$ & $\boldsymbol{22.28}$ & $9.42$ & $\boldsymbol{8.20}$ \\
			Ours & $\boldsymbol{1.29}$ & $\boldsymbol{5.07}$ & $11.54$ & $\boldsymbol{62.52}$ & $22.31$ & $\boldsymbol{9.25}$ & $8.25$ \\  \specialrule{1pt}{0pt}{0pt}
		\end{tabular}
\end{table} 
\end{scriptsize}
		
Apparently, our method of featurizing join predicates result in better Q-error performance, at the very slightly extra expense of memory cost and latency.

\vspace{-5pt}
\subsection{Effect of Number of Relations in Join}
\label{subsec:exp_case_studies}

We also investigate the effect of the number of relations involved in a query on~\modelName~and the alternatives. 
According to the number of relations involved, the testing sub-queries on STATS's \textsf{Insert-heavy} workload are divided into two categories. The sub-queries in both categories covers $\leq3$ and $\geq4$ join predicates, respectively.
Then, we observe each method's Q-error distributions on different categories. As the results in Fig.~\ref{fig:categ_times} show,~\modelName~is able to achieve the best overall performance. 
All methods have better Q-error distribution performance on queries involving less relations. However, the gap between~\modelName's achieved Q-error on queries with less and more relations is far smaller than that of the competitors. This implies that~\modelName~is effective at understanding the implicit relationships between true cardinalities and complex join patterns.
\begin{figure}[htb]
	\centering
	\if\includeAppendix1
	\includegraphics[width=0.45\textwidth, trim={5mm 2mm 1mm 9mm},clip]{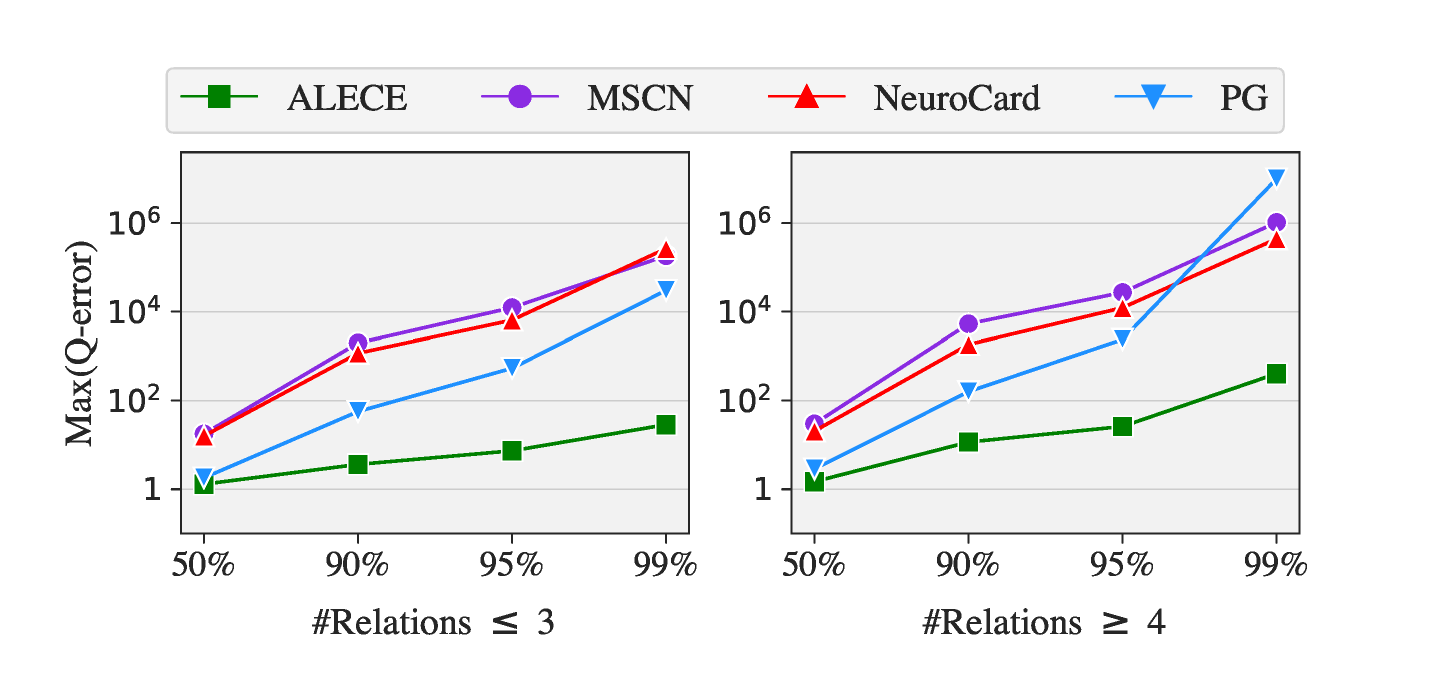}
	\else
	\includegraphics[width=0.45\textwidth, trim={5mm 2mm 1mm 9mm},clip]{categ_q_error.eps}
	\fi
	\vspace{-3pt}
	\caption{Q-error distributions with number of joins}
	\label{fig:categ_times}
\end{figure}

\subsection{Effect of DB State Types}
\label{subsec:exp_db_state_type}

We have conducted additional experiments to investigate the effects of the DB state types. We build three \modelName-variants with different types of the DB states:
\begin{itemize}[leftmargin=*]
	\item \textbf{\textsf{Histogram}} is of our used DB state type in the paper.
	\item \textbf{\textsf{Sample}} is built by uniformly sampling from the underlying data. For each relation $R$, we maintain a sample set $S_R$ of fixed size $M$. $S_R$ may gets updated when a data manipulation statement referred to the relation $R$ comes in. It is guaranteed that each tuple in $R$ is equal-possibly sampled. In our experiments, the value of $M$ is set to 512. In other words, the dimension of each DB state vector is 512. 
	
	\item \textbf{\textsf{NE-Hist}} is the set of histograms over unequally sized partitions over the value ranges of all attributes. Specifically, we first extract all values of an attribute $A$ from the initial dataset. Then the range of $A$ is partitioned into $\dx$ parts, with each part covering the same number of values. Each element of NE-Hist data featurzations is a vector generated by aligning the value frequency $f_i$ for each $i$, where $f_i$ refers to the number of values falling in the $i$th partition. It is noteworthy that the values of all $f_i$ are the same initially. With the data manipulation statements in the workload are executed, the value of $f_i$ will keep changing. 
\end{itemize}

Table~\ref{tab:exp_data_feat_type} shows the Q-error distributions, storage overhead and latency of three \modelName-variants with different types of the DB states on the \textsf{Insert-heavy} workload of the STATS datasets. The results on the other datasets and other workloads are similar and thus omitted.

\begin{scriptsize}
	\begin{table}[htb]
		\vspace{3pt}
		\centering
		\scriptsize
		\caption{The effects of DB state types}
		\label{tab:exp_data_feat_type}
		\vspace{-8pt}
		\arrayrulecolor{black}
			\begin{tabular}{  l V{2}   c  c  c  c | c | c }
				\specialrule{1pt}{0pt}{0pt}
				\multirow{2}{*} {DB state type} & \multicolumn{4}{c |}{Q-error} & \multirow{2}{*}{\tabincell{c}{Size \\ (MB)}} &  \multirow{2}{*}{\tabincell{c}{Latency \\ (ms)}} \\ \hhline{*{1}{|~}*{4}{|-}|} 
				& 50\% & 90\% & 95\% & 99\% &  &   \\ \specialrule{1pt}{0pt}{0pt}
				\textbf{\textsf{Histogram}} & $\boldsymbol{1.29}$ & $\boldsymbol{5.07}$ & $\boldsymbol{11.54}$ & $62.52$ & $\boldsymbol{22.31}$ & $8.25$ \\
				\textbf{\textsf{Sample}} & $1.52$ & $5.95$ & $12.86$ & $\boldsymbol{60.87}$ & $23.23$ & $8.67$\\
				\textbf{\textsf{NE-hist}} & $1.41$ & $6.01$ & $13.93$ & $77.74$ & $\boldsymbol{22.31}$ & $\boldsymbol{8.23}$ \\ \specialrule{1pt}{0pt}{0pt}
			\end{tabular}
		\end{table}
	\end{scriptsize}

Referring to Table~\ref{tab:exp_data_feat_type}, all of the three DB state types are able to help \modelName achieve accurate estimates. It is noteworthy that all of these features can be regarded as the marginal distribution approximation of single attributes. The data-encoder module of \modelName establishes a bridge between the marginal distributions and the joint distribution. It can quantitatively `calculate' the relevance between a pair of elements from any two DB states. Thus, it can effectively discover the implicit connections between any pair of attributes, which is helpful to the cardinality estimation task. 
	
Nevertheless, the cost of updating histograms when data manipulation statements come in is smaller than that of updating other types of DB states. Thus, our \modelName adopts the simple histograms as the DB states in the paper. In the future, we will investigate more types of DB states.

\fi
\end{document}